\documentclass[useamsfonts]{pasj01}

\usepackage{amssymb}
\usepackage{threeparttable}

\begin{document} 
\Received{mm/dd/2017}
\Accepted{mm/dd/2017}

\title{
SILVERRUSH. IV. Ly$\alpha$ Luminosity Functions \\
at $z = 5.7$ and $6.6$ Studied with 
$\sim$ 1,300
LAEs\\ 
on the $14-21$ deg$^2$ Sky
}

\author{Akira \textsc{Konno}\altaffilmark{1,2}}
\author{Masami \textsc{Ouchi}\altaffilmark{1,3}}
\author{Takatoshi \textsc{Shibuya}\altaffilmark{1}}
\author{Yoshiaki \textsc{Ono}\altaffilmark{1}}
\author{Kazuhiro \textsc{Shimasaku}\altaffilmark{2,4}}
\author{Yoshiaki \textsc{Taniguchi}\altaffilmark{5}}
\author{Tohru \textsc{Nagao}\altaffilmark{6}}
\author{Masakazu A. R. \textsc{Kobayashi}\altaffilmark{7}}
\author{Masaru \textsc{Kajisawa}\altaffilmark{6,8}}
\author{Nobunari \textsc{Kashikawa}\altaffilmark{9,10}}
\author{Akio K. \textsc{Inoue}\altaffilmark{11}}
\author{Masamune \textsc{Oguri}\altaffilmark{3,4,12}}
\author{Hisanori \textsc{Furusawa}\altaffilmark{9}}
\author{Tomotsugu \textsc{Goto}\altaffilmark{13}}
\author{Yuichi \textsc{Harikane}\altaffilmark{1,12}}
\author{Ryo \textsc{Higuchi}\altaffilmark{1,12}}
\author{Yutaka \textsc{Komiyama}\altaffilmark{9,10}}
\author{Haruka \textsc{Kusakabe}\altaffilmark{2}}
\author{Satoshi \textsc{Miyazaki}\altaffilmark{9,10}}
\author{Kimihiko \textsc{Nakajima}\altaffilmark{14}}
\author{Shiang-Yu \textsc{Wang}\altaffilmark{15}}

\altaffiltext{1}{Institute for Cosmic Ray Research, The University of Tokyo,
5-1-5 Kashiwanoha, Kashiwa, Chiba 277-8582, Japan}
\altaffiltext{2}{Department of Astronomy, Graduate School of Science,
The University of Tokyo, 7-3-1 Hongo, Bunkyo-ku, Tokyo 113-0033, Japan}
\altaffiltext{3}{Kavli Institute for the Physics and Mathematics
of the Universe (WPI), The University of Tokyo,
5-1-5 Kashiwanoha, Kashiwa, Chiba 277-8583, Japan}
\altaffiltext{4}{Research Center for the Early Universe, Graduate School of Science,
The University of Tokyo, 7-3-1 Hongo, Bunkyo, Tokyo 113-0033, Japan}
\altaffiltext{5}{The Open University of Japan, 2-11, Wakaba, Mihama-ku,
Chiba, Chiba 261-8586, Japan}
\altaffiltext{6}{Research Center for Space and Cosmic Evolution,Ehime University, 2-5 Bunkyo-cho,
Matsuyama, Ehime 790-8577, Japan}
\altaffiltext{7}{Faculty of Natural Sciences, National Institute of Technology, Kure College, 2-2-11
Agaminami, Kure, Hiroshima 737-8506, Japan}
\altaffiltext{8}{Graduate School of Science and Engineering, Ehime University, 2-5 Bunkyo-cho,
Matsuyama, Ehime 790-8577, Japan}
\altaffiltext{9}{National Astronomical Observatory of Japan,
2-21-1 Osawa, Mitaka, Tokyo 181-8588, Japan}
\altaffiltext{10}{SOKENDAI (The Graduate University for Advanced Studies),
2-21-1 Osawa, Mitaka, Tokyo 181-8588, Japan}
\altaffiltext{11}{
Department of Environmental Science and Technology, Faculty of Design Technology, Osaka Sangyo University, 
3-1-1 Nakagaito, Daito, Osaka 574-8530, Japan}
\altaffiltext{12}{Department of Physics, Graduate School of Science,
The University of Tokyo, 7-3-1 Hongo, Bunkyo-ku, Tokyo 113-0033, Japan}
\altaffiltext{13}{Institute of Astronomy, National Tsing Hua University, No. 101,
Section 2, Kuang-Fu Road, Hsinchu, Taiwan}
\altaffiltext{14}{European Southern Observatory, Karl-Schwarzschild-Str. 2, D-85748
Garching bei Munchen, Germany}
\altaffiltext{15}{Academia Sinica, Institute of Astronomy and Astrophysics,
No.1, Sec. 4, Roosevelt Rd, Taipei 10617, Taiwan}

%



\maketitle

\begin{abstract}
We present the Ly$\alpha$ luminosity functions (LFs) at $z = 5.7$ and $6.6$ derived 
from a new large sample of 
1,266 Ly$\alpha$ emitters (LAEs) identified in total areas of $14$ and $21$ deg$^2$, respectively,
based on the early narrowband data of the Subaru/Hyper Suprime-Cam (HSC) survey. 
Together with careful Monte-Carlo simulations that account for the incompleteness of the LAE 
selection and the flux estimate systematics in the narrowband imaging, we have determined
the Ly$\alpha$ LFs with the unprecedentedly small statistical and systematic uncertainties 
in a wide Ly$\alpha$ luminosity range of $10^{42.8-43.8}$ erg s$^{-1}$. 
We obtain the best-fit Schechter parameters of 
$L^{*}_{\mathrm{Ly}\alpha} = 1.6^{+2.2}_{-0.6} \ (1.7^{+0.3}_{-0.7}) \times 10^{43} \ \mathrm{erg} \ \mathrm{s}^{-1}$, 
$\phi^{*}_{\mathrm{Ly}\alpha} = 0.85^{+1.87}_{-0.77} \ (0.47^{+1.44}_{-0.44}) \times 10^{-4} \ \mathrm{Mpc}^{-3}$, 
and 
$\alpha = -2.6^{+0.6}_{-0.4} \ (-2.5^{+0.5}_{-0.5})$ 
at $z=5.7$ ($6.6$).
We confirm that our best-estimate Ly$\alpha$ LFs are consistent with the majority of the previous studies, 
but find that our Ly$\alpha$ LFs 
do not agree with the high number densities of LAEs 
recently claimed by Matthee/Santos et al.'s  
studies that may overcorrect the incompleteness and the flux systematics.
Our Ly$\alpha$ LFs at $z=5.7$ and $6.6$ show 
an indication that 
the faint-end slope is very steep ($\alpha \simeq -2.5$), 
although it is also possible that 
the bright-end LF results are enhanced by systematic effects 
such as 
the contribution from AGNs, 
blended merging galaxies, 
and/or 
large ionized bubbles around bright LAEs. 
Comparing our Ly$\alpha$ LF measurements
with four independent reionization models,
we estimate the neutral hydrogen fraction of the IGM to be 
$x_\mathrm{HI} = 0.3 \pm 0.2$ 
at $z = 6.6$
that is consistent with the small Thomson scattering optical
depth obtained by \textit{Planck} 2016.
\end{abstract}

\section{Introduction}

Ly$\alpha$ emission lines are one of the key properties of galaxies
for exploring a high-$z$ universe.
Ly$\alpha$ emitters (LAEs), which generally have a spectrum of a luminous Ly$\alpha$ line
and a faint ultraviolet (UV) continuum, have been found at a wide redshift range of 
$z = 0 - 8$ by 
several approaches including narrowband surveys 
(e.g., \cite{1998AJ....115.1319C}; \cite{1998ApJ...502L..99H}; \cite{2000ApJ...545L..85R}; 
\cite{2000ApJ...532..170S}; \cite{2002ApJ...565L..71M}; \cite{2002ApJ...576L..25A}; 
\cite{2003ApJ...582...60O}; 
\cite{2004AJ....128.2073H}; \cite{2004AJ....128..569M}; \cite{2005PASJ...57..165T}; 
\cite{2006Natur.443..186I}; \cite{2006ApJ...648....7K}; \cite{2006PASJ...58..313S}; 
\cite{2007ApJ...667...79G}; \cite{2007ApJS..172..523M}; \cite{2010ApJ...714..255G}; 
\cite{2012ApJ...752..114S}; \cite{2012AJ....143...79Y}; \cite{2014ApJ...797...16K}) 
and spectroscopic observations 
(e.g., \cite{2008ApJ...680.1072D}; \cite{2011ApJS..192....5A}; \cite{2013Natur.502..524F}; 
\cite{2014ApJ...795...20S}; \cite{2015A&A...573A..24C}; \cite{2015ApJ...804L..30O}; 
\cite{2015ApJ...810L..12Z}; \cite{2016ApJ...826..113S}; \cite{2017MNRAS.464..469S}). 
From these observations, it has been revealed that LAEs are in an early phase of
galaxy evolution, i.e., LAEs are young, less massive, less dusty, and in highly ionized state
(e.g., \cite{2010MNRAS.402.1580O}; \cite{2010ApJ...724.1524O}; \cite{2014MNRAS.442..900N}; 
\cite{2015ApJ...800L..29K}; \cite{2016Sci...352.1559I}).

Ly$\alpha$ luminosity functions (LFs) and their evolution
can be a probe for the early evolution of galaxies and cosmic reionization 
(e.g., \cite{1999ApJ...518..138H}; 
\cite{2007MNRAS.381...75M}; 
\cite{2007ApJ...667..655M}; 
\cite{2007ApJ...670..919K}; 
\cite{2008MNRAS.386.1990M}; 
\cite{2011MNRAS.410..830D}).
Previous studies have found that Ly$\alpha$ LFs increase from $z \sim 0$ to $z \sim 3$,
show a moderate plateau between $z \sim 3$ to $z \sim 6$,
and decrease toward $z \gtrsim 6$
(e.g., \cite{2008ApJ...680.1072D}; \cite{2008ApJS..176..301O}; \cite{2011ApJ...734..119K}). 
The evolution of Ly$\alpha$ LFs is different from that of UV LFs,
which increases from $z \sim 0$ to $z \sim 2$, and turns to the decrease beyond $z \gtrsim 3$
(e.g., \cite{2005ApJ...619L..47S}; \cite{2009ApJ...692..778R}; \cite{2015ApJ...803...34B}; 
see also Figure 7 of \cite{2016ApJ...823...20K}). 
The difference of the evolutionary trend between Ly$\alpha$ and UV LFs would be related to
the escaping process of Ly$\alpha$ photons not only from the \textsc{Hi} ISM
of a galaxy, but also from the \textsc{Hi} intergalactic medium (IGM).
The Ly$\alpha$ escape fraction, $f^{\mathrm{Ly}\alpha}_\mathrm{esc}$,
which is defined by the ratio of the star formation rate densities (SFRDs)
estimated from observed Ly$\alpha$ luminosity densities (LDs)
to those estimated from intrinsic UV LDs,
largely increases from $z \sim 0$ to $z \sim 6$ by two orders of magnitudes,
and turns to the decrease beyond $z \gtrsim 6$
(e.g., \cite{2011ApJ...730....8H}). 
The rapid evolution of the Ly$\alpha$ escape fraction from $z \sim 6$ to $z \sim 0$ 
would be explained by
the combination of the Ly$\alpha$ attenuation by dust
and the Ly$\alpha$ resonance scattering effect by \textsc{Hi} in ISM.
In the case that the ISM \textsc{Hi} density of a galaxy is large,
the path lengths of Ly$\alpha$ photons become longer due to the resonant scattering, and
these Ly$\alpha$ photons are subject to the attenuation by dust.
\citet{2016ApJ...823...20K} have used simple expanding shell models, which compute the Ly$\alpha$ radiative transfer
by Monte Carlo simulations 
(MCLya; \cite{2006A&A...460..397V}; \cite{2011A&A...531A..12S}),
and have suggested that the large increase of Ly$\alpha$ escape fraction at $z = 0-6$
can be reproduced by the combination of the \textsc{Hi} column density decrease
(by two orders of magnitude) and the average dust extinction values.
The decrease of the Ly$\alpha$ LFs at $z \gtrsim 6$ is related to the cosmic reionization,
because the Ly$\alpha$ damping wing of \textsc{Hi} in IGM attenuates Ly$\alpha$ photons from a galaxy.
Previous studies have found that Ly$\alpha$ LFs at $z \sim 7$ significantly
decrease from those at $z \sim 6$ 
(e.g., \cite{2006ApJ...648....7K}; \cite{2010ApJ...723..869O}; \cite{2010ApJ...725..394H}; \cite{2016MNRAS.463.1678S}), 
and especially at $z \gtrsim 7$, Ly$\alpha$ LFs decrease rapidly 
(e.g., \cite{2014ApJ...797...16K}). 
The neutral hydrogen fraction of IGM, $x_\mathrm{HI}$, can be estimated 
by the Ly$\alpha$ LD evolution subtracting the galaxy evolution effect.
\citet{2010ApJ...723..869O}
have constrained $x_\mathrm{HI} = 0.2 \pm 0.2$ at $z = 6.6$ from the Ly$\alpha$ LF
evolution at $z = 5.7 - 6.6$ 
(see also \cite{2004ApJ...617L...5M}; \cite{2006ApJ...648....7K}). 
Similarly, the neutral hydrogen fractions at $z \gtrsim 7$ have also been estimated
from the Ly$\alpha$ LF evolution 
(\cite{2010ApJ...722..803O}; \cite{2014ApJ...797...16K}; 
\cite{2017ApJ...844...85O}). 
These  $x_\mathrm{HI}$ estimates could constrain the history of cosmic reionization
by the comparison with the Thomson scattering optical depth
of cosmic microwave background (CMB).

Recently, a large number of wide-field narrowband imaging surveys have been conducted
not only to spread the Ly$\alpha$ luminosity ranges of Ly$\alpha$ LFs,
but also to reveal physical properties for luminous LAEs.
At $z \sim 2-3$, luminous LAEs are known to have counterparts
in multiwavelength data (e.g., X-ray and radio)
and/or extended Ly$\alpha$ haloes
(e.g., \cite{2000ApJ...532..170S}; \cite{2008ApJS..176..301O}; \cite{2014Natur.506...63C}; \cite{2017ApJ...837...71C}). 
A recent study, for example, has confirmed that there are excesses
found in Ly$\alpha$ LFs at $\log L(\mathrm{Ly}\alpha)$ [erg s$^{-1}$] $\gtrsim 43.4$,
and the excesses are made by (faint) AGNs based on multiwavelength imaging data 
\citep{2016ApJ...823...20K}. 
Interestingly, such luminous LAEs have also been discovered at a higher redshift of $z \sim 6.6$
(e.g., Himiko by \cite{2009ApJ...696.1164O}, 
CR7 and MASOSA by \cite{2015ApJ...808..139S}, 
and COLA1 by \cite{2016ApJ...825L...7H};
see also IOK-1 by \cite{2006Natur.443..186I}).
A number of observational and theoretical studies have aimed to uncover
the physical origins of these bright LAEs 
(e.g., \cite{2013ApJ...778..102O} and \cite{2015MNRAS.451.2050Z} for Himiko; 
\cite{2017MNRAS.469..448B}, \cite{2017MNRAS.468L..77P}, and \cite{2017arXiv170500733S} for CR7).

\begin{figure}
\begin{center}
\includegraphics[width=8cm]{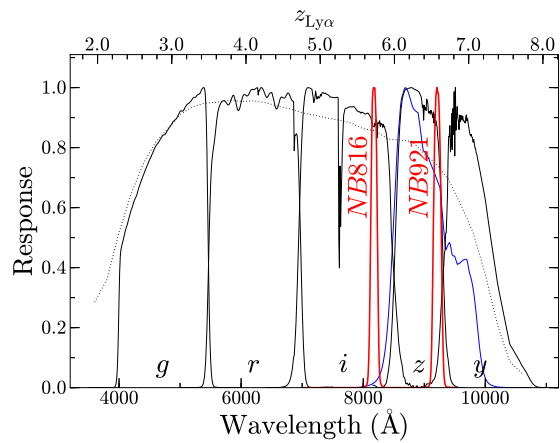}
\end{center}
\caption{
Filter response curves for the broadband and narrowband filters of Subaru/HSC.
The red lines at the wavelength of $\sim 8100$\AA\ and $\sim 9200$\AA\
are the transmission curves of \textit{NB816} and \textit{NB921}, respectively. 
The black solid curves denote the curves for the broadband filters
(\textit{g}, \textit{r}, \textit{i}, \textit{z}, and \textit{y}). 
These response curves 
take account of 
the CCD quantum efficiency of HSC (black dotted line),
airmass, 
the transmittance of the dewar window and primary focus unit, 
and the reflectivity of the primary mirror. 
For reference, we also plot 
the filter response curve of the $z'$-band filter of Subaru/Suprime-Cam 
(blue solid curve). 
The peaks of these curves are normalized to 1.0 for clarity. 
The upper $x$-axis shows the redshift of Ly$\alpha$.
}\label{fig:HSCfilter}
\end{figure}

\begin{table*}
\caption{
Summary of HSC/\textit{NB816} and \textit{NB921} Data}
\begin{center}
\begin{threeparttable}
\small
\begin{tabular}{ccccccccccc}
\hline
\hline
Field & Area (\textit{NB816}) & Area (\textit{NB921}) &$g$\tnote{a} & $r$\tnote{a} & $i$\tnote{a} & $\textit{NB816}$\tnote{a} & $z$\tnote{a} & $\textit{NB921}$\tnote{a} & $y$\tnote{a} \\
 & (deg$^2$) & (deg$^2$) & (ABmag) & (ABmag) & (ABmag) & (ABmag) & (ABmag) & (ABmag) & (ABmag)\\
\hline
UD-COSMOS & 1.97 & 2.05 & 26.9 & 26.6 & 26.2 & 25.7 & 25.8 & 25.6 & 25.1 \\
UD-SXDS & 1.93 & 2.02 & 26.9 & 26.4 & 26.3 & 25.5 & 25.6 & 25.5 & 24.9 \\
D-COSMOS & --- & 5.31 & 26.5 & 26.1 & 26.0 & --- & 25.5 & 25.3 & 24.7 \\
D-DEEP2-3 & 4.37 & 5.76 & 26.6 & 26.2 & 25.9 & 25.2 & 25.2 & 24.9 & 24.5 \\
D-ELAIS-N1 & 5.56 & 6.08 & 26.7 & 26.0 & 25.7 & 25.3 & 25.0 & 25.3 & 24.1 \\
\hline 
Total & 13.8 & 21.2 & --- & --- & --- & --- & --- & --- & --- \\
\hline 
\end{tabular}\label{table:imaging_data_HSC}
\begin{tabnote}
The narrowband and broadband data are obtained in the HSC SSP survey.
\item[a] The $5\sigma$ limiting magnitude in a circular aperture with a diameter of $1\farcs5$.
\end{tabnote}
\end{threeparttable}
\end{center}
\end{table*}

In this paper, we present the Ly$\alpha$ LFs at $z = 5.7$ and $6.6$
based on the Subaru/Hyper Suprime-Cam (HSC) Subaru Strategic Program (SSP; \cite{2017arXiv170405858A}). 
Because the field of view of HSC is about seven times wider than that of Subaru/Suprime-Cam,
HSC can identify a large number of high-$z$ LAEs with a wide range of
Ly$\alpha$ luminosity more efficiently than Suprime-Cam.
In our HSC SSP survey, 
a total of $\sim 13.8$ deg$^2$ and $\sim 21.2$ deg$^2$ sky areas 
are covered by \textit{NB816} and \textit{NB921} observations, respectively 
(see also Section \ref{sec:sample_image}, \cite{2017arXiv170407455O} and \cite{2017arXiv170408140S} for details). 
These wide field HSC \textit{NB} data sets 
allow us to determine the Ly$\alpha$ LFs at $z = 5.7$ and $6.6$
with unprecedented accuracy.
By examining the evolution of these Ly$\alpha$ LFs at $z = 5.7 - 6.6$,
we can constrain the $x_\mathrm{HI}$ value at $z = 6.6$.
Moreover, based on these HSC SSP data, 
we can push the Ly$\alpha$ luminosity range toward brighter luminosity, 
and investigate the abundance of luminous high-$z$ LAEs. 
We describe a summary of our HSC surveys 
and the sample construction for $z = 5.7$ and $6.6$ LAEs in Section \ref{sec:sample}.
We derive the Ly$\alpha$ LFs at these redshifts,
and compare the Ly$\alpha$ LFs with those of previous studies in Section \ref{sec:LF}.
We examine the Ly$\alpha$ LF evolution at $z = 5.7 - 6.6$,
and discuss cosmic reionization in Section \ref{sec:discuss}.
This paper is placed in a series of papers from twin programs studying high-$z$ objects
based on the HSC SSP data products.
One program is our high-$z$ LAE studies 
named Systematic Identification of LAEs for Visible Exploration and Reionization Research 
Using Subaru HSC (SILVERRUSH). 
This program provides the clustering measurements of $z=5.7$ and $6.6$ LAEs
\citep{2017arXiv170407455O}, 
the photometric and spectroscopic properties of LAEs at these redshifts
\citep{2017arXiv170408140S,2017arXiv170500733S}, 
the systematic survey for LAE overdense region
(R. Higuchi et al. in preparation),
and our Ly$\alpha$ LF studies.
The other program is the high-$z$ dropout galaxy study, 
Great Optically Luminous Dropout Research Using Subaru HSC 
(GOLDRUSH; \cite{2017arXiv170406004O}; \cite{2017arXiv170406535H}; 
\cite{2017arXiv170809421T}). 
Throughout this paper, we use magnitudes in the AB system 
\citep{1974ApJS...27...21O}. 
We adopt $\Lambda$CDM cosmology with a parameter set of
($h$, $\Omega_m$, $\Omega_\Lambda$, $\sigma_8$) = ($0.7$, $0.3$, $0.7$, $0.8$),
which is consistent with the nine-year \textit{WMAP} and the latest \textit{Planck} results
\citep{2013ApJS..208...19H,2016A&A...594A..13P}.

\section{Observations and Sample Selection}
\label{sec:sample}

\subsection{Hyper Suprime-Cam Imaging Observations and\\
Data Reduction}
\label{sec:sample_image}

In our sample construction for $z = 5.7$ and $6.6$ LAEs,
we use narrowband (\textit{NB816}, \textit{NB921}) imaging data
as well as broadband ($g, r, i, z, y$) imaging data,
which are taken with Subaru/HSC 
(\cite{2012SPIE.8446E..0ZM}; 
see also \cite{miyazaki2017}; \cite{furusawa2017}; 
\cite{kawanomoto2017}; \cite{komiyama2017}). 
The narrowband filters, \textit{NB816} and \textit{NB921}, have central wavelengths of 8170 \AA\ and 9210 \AA, respectively, 
and FWHMs of 131 \AA\ and 120 \AA\ to identify LAEs in the redshift range of 
$z = 5.67 - 5.77$ and $z = 6.52 - 6.63$, respectively.
We show the response curves of the narrowband filters as well as the broadband filters
in Figure \ref{fig:HSCfilter}.
These narrowband and broadband images are obtained
in our ongoing HSC legacy survey under the Subaru Strategic Program
(SSP; PI: S.Miyazaki, see also \cite{2017arXiv170405858A}).
The HSC SSP has been allocated 300 nights over 5 years, and started in March 2014.
The HSC SSP survey has three layers with different sets of area and depth: 
the Wide, Deep, and UltraDeep layers.
These layers will cover the sky area of $\sim 1400$ deg$^2$, $\sim 30$ deg$^2$, and $\sim 4$ deg$^2$
with the $5\sigma$ limiting magnitudes (in \textit{r} band)
of $\sim 26$ mag, $\sim 27$ mag, and $\sim 28$ mag, respectively.
While the broadband images are taken in all the three layers,
the \textit{NB816} and \textit{NB921} images are obtained only in the Deep and UltraDeep layers.
We use early datasets of the HSC SSP survey taken from March 2014 to April 2016 (S16A), 
where all additional data taken in January to April 2016 have been merged 
with the data of Public Data Release 1 \citep{2017arXiv170208449A}.  
With the \textit{NB816} filter, 
the HSC SSP survey has observed two blank fields in the Deep layer,
the D-DEEP2-3 ($23^\mathrm{h} 30^\mathrm{m} 00^\mathrm{s}$, $+00^\mathrm{d} 00' 00\farcs0$)
and D-ELAIS-N1 ($16^\mathrm{h} 10^\mathrm{m} 00^\mathrm{s}$, $+54^\mathrm{d} 00' 00\farcs0$)
fields, and two blank fields in the UltraDeep layer,
the UD-COSMOS ($10^\mathrm{h} 00^\mathrm{m} 29^\mathrm{s}$, $+02^\mathrm{d} 12' 21\farcs0$)
and UD-SXDS ($02^\mathrm{h} 18^\mathrm{m} 00^\mathrm{s}$, 
$-05^\mathrm{d} 00' 00\farcs0$) fields.
For the \textit{NB921} filter, 
a blank field of
the D-COSMOS ($10^\mathrm{h} 00^\mathrm{m} 29^\mathrm{s}$, $+02^\mathrm{d} 12' 21\farcs0$) field
in the Deep layer has also been observed as well as the four fields described above.
Each field in the Deep layer is covered by three or four pointing positions of HSC,
while in the UltraDeep layer,
each field is covered by one pointing position of HSC.
The details of our HSC SSP survey is listed in Table \ref{table:imaging_data_HSC}.

The HSC data are reduced by the HSC SSP survey team with \texttt{hscPipe}
\citep{2017arXiv170506766B}, 
which is based on the Large Synoptic Survey Telescope
(LSST) pipeline 
\citep{2008arXiv0805.2366I,2010SPIE.7740E..15A,2015arXiv151207914J}. 
This HSC pipeline performs CCD-by-CCD reduction,
calibrates astrometry, mosaic-stacking, and photometric zeropoints,
and generates catalogs for sources detected and photometrically measured in the stacked images.
The photometric and astrometric calibrations are based on the data from
the Panoramic Survey Telescope and Rapid Response System 1 imaging survey
(Pan-STARRS1; \cite{2012ApJ...756..158S,2012ApJ...750...99T,2013ApJS..205...20M}).
In the stacked images, regions contaminated with diffraction spikes
and halos of bright stars are masked by using the mask extension outputs of the HSC pipeline 
\citep{2017arXiv170500622C}.
After the masking, the total effective survey areas in the S16A data are $13.8$ deg$^2$ and $21.2$ deg$^2$
for \textit{NB816} and \textit{NB921}, respectively.
These survey areas are $70-87$ times larger than 
those of the Subaru Deep Field studies (\cite{2006PASJ...58..313S}; \cite{2011ApJ...734..119K}), 
$14-21$ times larger than 
those of the Subaru/\textit{XMM-Newton} Deep Survey (\cite{2008ApJS..176..301O}; \cite{2010ApJ...723..869O}), 
and $2-5$ times larger than 
those of other subsequent studies with Subaru/Suprime-Cam 
(\cite{2015MNRAS.451..400M}; \cite{2016MNRAS.463.1678S}). 
Under the assumption of a simple top-hat selection function for LAEs whose redshift distribution is defined
by the FWHM of a narrowband filter, these survey areas correspond to comoving volumes of
$\simeq 1.16 \times 10^7$ Mpc$^3$ and $\simeq 1.91 \times 10^7$ Mpc$^3$
for $z=5.7$ and $6.6$ LAEs, respectively.
The narrowband images reach the $5\sigma$ limiting magnitudes in a $1\farcs5$-diameter
circular aperture of $24.9 - 25.3$ mag in the Deep layer, and $25.5 - 25.7$ mag in the UltraDeep layer.
Note that the PSF sizes of the HSC images are typically $< 0\farcs8$,
which is sufficiently smaller than the aperture diameter of $1\farcs5$
(see \cite{2017arXiv170208449A} for details). 
We summarize 
the $5\sigma$ limiting magnitudes
of the \textit{NB816} and \textit{NB921} images in Table \ref{table:imaging_data_HSC}.
For the total magnitudes, 
we use cmodel magnitudes. 
The cmodel magnitude is derived from 
a linear combination of exponential and de Vaucouleurs profile fits 
to the light profile of each object 
\citep{2017arXiv170506766B}. 
We make use of the cmodel magnitudes for color measurements, 
because the HSC data used in this study are reduced 
with no smoothing to equalize the PSFs 
and fixed aperture photometry does not provide 
good measurements of object colors 
\citep{2017arXiv170208449A}.  
The total magnitudes and colors are corrected for Galactic extinction 
\citep{1998ApJ...500..525S}.

\subsection{Photometric Samples of $z=5.7$ and $6.6$ LAEs}
\label{sec:sample_sample}

\begin{table}
\caption{
Photometric Sample of $z = 5.7$ and $6.6$ LAEs}
\begin{center}
\begin{threeparttable}
\small
\begin{tabular}{lcc}
\hline
\hline
Field 	& LAE All sample\tnote{a}	& LAE Ly$\alpha$ LF sample\tnote{b}	\\
\hline
\multicolumn{3}{c}{The $z = 5.7$ LAE sample}	\\
\hline
UD-COSMOS			& 201	& 201					\\
UD-SXDS				& 224	& 224					\\
D-DEEP2-3			& 423	& 423					\\
D-ELAIS-N1			& 229	& 229					\\
\hline
Total					& 1077	& 1077 					\\
\hline
\hline
\multicolumn{3}{c}{The $z = 6.6$ LAE sample}	\\
\hline
UD-COSMOS			& 338	& 50					\\
UD-SXDS				& 58		& 21					\\
D-COSMOS			& 244	& 48					\\
D-DEEP2-3			& 164	& 38					\\
D-ELAIS-N1			& 349	& 32					\\
\hline
Total					& 1153	& 189		\\		
\hline
\end{tabular}\label{table:sample_HSC}
\begin{tabnote}
\item[a] The numbers of LAE candidates 
selected based on the color selection criteria (Equations \ref{eq:5p7select} and \ref{eq:6p6select}) 
and the contamination rejection process \citep{2017arXiv170408140S}. 
\item[b] The numbers of LAE candidates used in our Ly$\alpha$ LF measurements. 
For the $z=6.6$ sample, we adopt a more stringent $z -$\textit{NB921} color criterion  
(Section \ref{sec:sample_sample}).
\end{tabnote}
\end{threeparttable}
\end{center}
\end{table}

LAE samples at $z = 5.7$ and $6.6$ are constructed 
based on narrowband color excess by Ly$\alpha$ emission,
$i - \textit{NB816}$ and $z - \textit{NB921}$, respectively, 
and no detection of blue continuum fluxes. 
We first select objects with magnitudes brighter than 
the $5\sigma$ limit in \textit{NB816} or \textit{NB921} 
from the HSC SSP database. 
We then apply similar selection criteria to those of 
\citet{2008ApJS..176..301O} and \citet{2010ApJ...723..869O}:  
\begin{equation}
\begin{array}{l}
i - \textit{NB816} \geq 1.2, \\
g > g_{3 \sigma}, \\
\mathrm{and} \ [ ( r \leq r_{3 \sigma} \ \mathrm{and} \ r - i \geq 1.0) \ \mathrm{or} \ (r > r_{3 \sigma}) ] 
\end{array}
\label{eq:5p7select}
\end{equation}
for $z = 5.7$ LAEs, and
\begin{equation}
\begin{array}{l}
z - \textit{NB921} \geq 1.0, \\
g > g_{3 \sigma}, \\
r > r_{3 \sigma}, \\
\mathrm{and} \ [ ( z \leq z_{3 \sigma} \ \mathrm{and} \ i - z \geq 1.0) \ \mathrm{or} \ (z > z_{3 \sigma}) ] 
\end{array}
\label{eq:6p6select}
\end{equation}
for $z = 6.6$ LAEs, 
where ($g_{3 \sigma}$, $r_{3 \sigma}$, $z_{3 \sigma}$) 
are the $3 \sigma$ limiting magnitudes of ($g$, $r$, $z$) bands. 
Note that 
the criterion in the former parentheses 
of the third criterion in Equation (\ref{eq:5p7select}) and 
the fourth criterion in Equation (\ref{eq:6p6select}) are 
used to select bright objects whose SED is consistent with 
a Lyman break due to intergalactic absorption.
In addition to the color selection criteria, we use the \texttt{countinputs} parameter,
which represents the number of exposures for each object in each band.
We apply \texttt{countinputs} $\ge 3$ for the narrowband images.
We also remove objects affected by bad pixels, proximity to bright stars,
or poor photometric measurement 
by using the following flags: 
\texttt{flags\_pixel\_edge}, \texttt{flags\_pixel\_interpolated\_center}, \texttt{flags\_pixel\_saturated\_center},
\texttt{flags\_pixel\_cr\_center}, and \texttt{flags\_pixel\_bad}.
After the visual inspection for the rejection of spurious sources and cosmic rays,
we identify 1,081 and 1,273 LAE candidates at $z = 5.7$ and $6.6$, respectively 
\citep{2017arXiv170408140S}. 
The samples of these LAE candidates are referred to as the {\lq}LAE All{\rq} samples.
The LAE All samples are $\sim 2-6$ times larger than
photometric samples in previous studies 
(e.g., \cite{2008ApJS..176..301O}; \cite{2010ApJ...723..869O}; \cite{2015MNRAS.451..400M}; \cite{2016MNRAS.463.1678S}). 
This sample is used for clustering analyses in our companion paper \citep{2017arXiv170407455O}. 
The details of the sample construction 
including the color-magnitude diagrams of \textit{NB}$-$\textit{BB} vs. \textit{NB} are 
presented in \citet{2017arXiv170408140S}.

In this Ly$\alpha$ LF study, 
we create subsamples of the LAE All samples 
to directly compare our results with previous work. 
The only difference between the subsamples and the LAE All samples 
is the $z -$\textit{NB921} color criterion for $z=6.6$ LAEs. 
The color selection criterion for $z = 5.7$ LAEs
(i.e., $\textit{i} - \textit{NB816} > 1.2$ in Equation \ref{eq:5p7select})
corresponds to the rest-frame Ly$\alpha$ equivalent width (EW), 
EW$_0$, of $\mathrm{EW}_0 \gtrsim 10$\AA\
in the case of a flat UV continuum (i.e., $f_{\nu} =$ const.) with IGM attenuation  
\citep{1995ApJ...441...18M}. 
This EW limit is similar to those of previous studies
($\mathrm{EW}_0 \gtrsim 10 - 30$\AA; e.g., 
\cite{2006PASJ...58..313S}; \cite{2008ApJS..176..301O}; \cite{2016MNRAS.463.1678S}). 
Thus, the $z=5.7$ LAE sample of the LAE All samples can be used for comparison 
with the previous Ly$\alpha$ LF results. 
On the other hand, 
the color criterion of  $z - \textit{NB921} > 1.0$
in Equation (\ref{eq:6p6select}) for $z = 6.6$ LAEs
corresponds to the EW$_0$ limit significantly lower than those of previous studies using Subaru/Suprime-Cam
(e.g., \cite{2010ApJ...723..869O}; \cite{2015MNRAS.451..400M}). 
This is because 
the relative wavelength position of \textit{NB921} to $z'$ (or $z$) band
filter is different between Suprime-Cam and HSC 
(Figure \ref{fig:HSCfilter}). 
Specifically, 
the central wavelength of the HSC $z$-band filter is 
$\simeq 160${\AA} shorter than that of the Suprime-Cam $z'$-band filter. 
For consistency of comparison, 
we adopt a more stringent color criterion of $z - \textit{NB921} > 1.8$. 
This criterion corresponds to EW$_0 > 14$\AA\ ($f_{\nu} =$ const.),
which is the same as that used in \citet{2010ApJ...723..869O}. 
We refer to these $z=5.7$ and $6.6$ LAE samples as the {\lq}LAE Ly$\alpha$ LF{\rq} samples. 
We use the LAE Ly$\alpha$ LF samples to derive surface number densities and color distributions 
(Section \ref{sec:LF_snd_cd}), and Ly$\alpha$ LFs at $z = 5.7$ and $6.6$ (Section \ref{sec:LF_z5p76p6}).
The numbers of our LAE candidates at $z = 5.7$ and $6.6$ are summarized in Table \ref{table:sample_HSC}. 
Note that the number of $z=5.7$ LAEs found in D-DEEP2-3 
is about two times larger than that in D-ELAIS-N1, 
although the area of D-DEEP2-3 is about 1.3 times smaller than that of D-ELAIS-N1 
and the depths of the \textit{NB816} data for these two fields are comparable. 
This is probably because the seeing of the \textit{NB816} data 
for D-DEEP2-3 is better than that for D-ELAIS-N1. 
This is also the case for the difference of the numbers 
of $z=6.6$ LAEs between UD-COSMOS and UD-SXDS.

\section{Ly$\alpha$ Luminosity Functions}
\label{sec:LF}

\subsection{Detection Completeness}
\label{sec:LF_detcomp}

\begin{figure}
\begin{center}
\includegraphics[width=8cm]{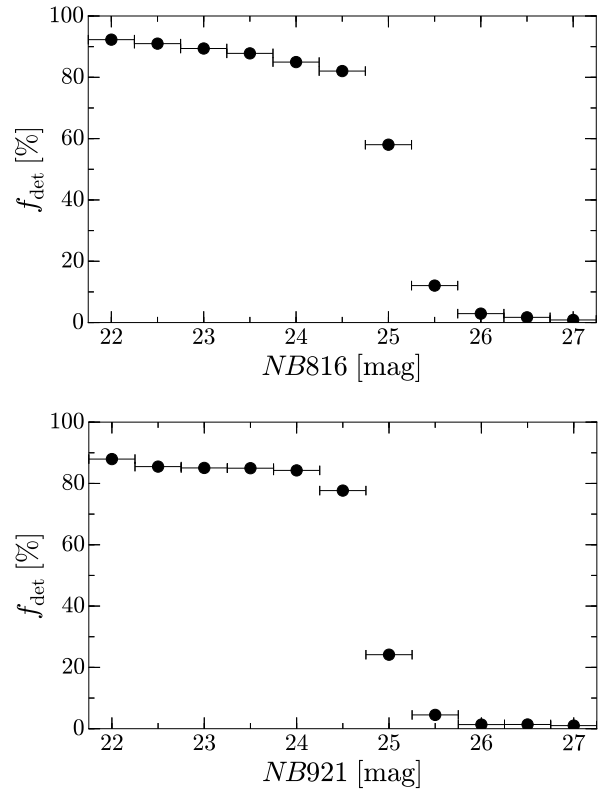}
\end{center}
\caption{Detection completeness, $f_\mathrm{det}$, of
our \textit{NB816} and \textit{NB921} images taken with Subaru/HSC.
The filled circles in the top and bottom figures represent the completeness values 
in a magnitude bin of $\Delta m = 0.5$ mag 
as a function of narrowband magnitude 
in the D-DEEP2-3 field. 
The $5\sigma$ limiting magnitudes of the \textit{NB816} and \textit{NB921} images
are 25.2 mag and 24.9 mag, respectively. 
}\label{fig:det_comp_hsc}
\end{figure}

We estimate detection completeness as a function of
the \textit{NB816} and \textit{NB921} magnitude by Monte Carlo simulations
with the \texttt{SynPipe} software 
\citep{2017arXiv170501599H,murata2017}. 
Using the \texttt{SynPipe} software, we distribute $\sim$ 18,000 pseudo LAEs
with various magnitudes in \textit{NB816} and \textit{NB921} images.
These pseudo LAEs have a S\'{e}rsic profile with the S\'{e}rsic index of $n=1.5$,
and the half-light radius of $r_\mathrm{e} \sim 0.9$ kpc, 
which corresponds to $0.15$ and $0.17$ arcsec for $z=5.7$ and $6.6$ sources, respectively.  
These S\'{e}rsic index and half-light radius values are similar to the average ones of $z \sim 6$ LBGs
with $L_\mathrm{UV} = 0.3 - 1 L^{*}_{z = 3}$ \citep{2015ApJS..219...15S}. 
We then perform source detection and photometry with \texttt{hscPipe},  
and calculate the detection completeness.
We define the detection completeness in a magnitude bin 
as the fraction of the numbers of the detected pseudo LAEs
to all of the input pseudo LAEs in the magnitude bin.
Figure \ref{fig:det_comp_hsc} shows the detection completeness
of the \textit{NB816} and \textit{NB921} images for the D-DEEP2-3 field. 
We find that the detection completeness is typically $\gtrsim 80$\%\ 
for bright objects with $\textit{NB} \lesssim 24.5$ mag, and 
$\sim 40$\%\ at the $5\sigma$ limiting magnitudes of these narrowband images. 
We correct for the detection completeness  
to derive the surface number densities and the Ly$\alpha$ LFs of LAEs 
in Sections \ref{sec:LF_snd_cd} and \ref{sec:LF_z5p76p6}.  
For the D-DEEP2-3 field, 
we use the detection completeness shown in Figure \ref{fig:det_comp_hsc}, 
and 
for the other fields, 
we shift it along the magnitude considering the limiting magnitudes of the narrowband images.

\begin{figure*}
\begin{center}
\includegraphics[width=11cm]{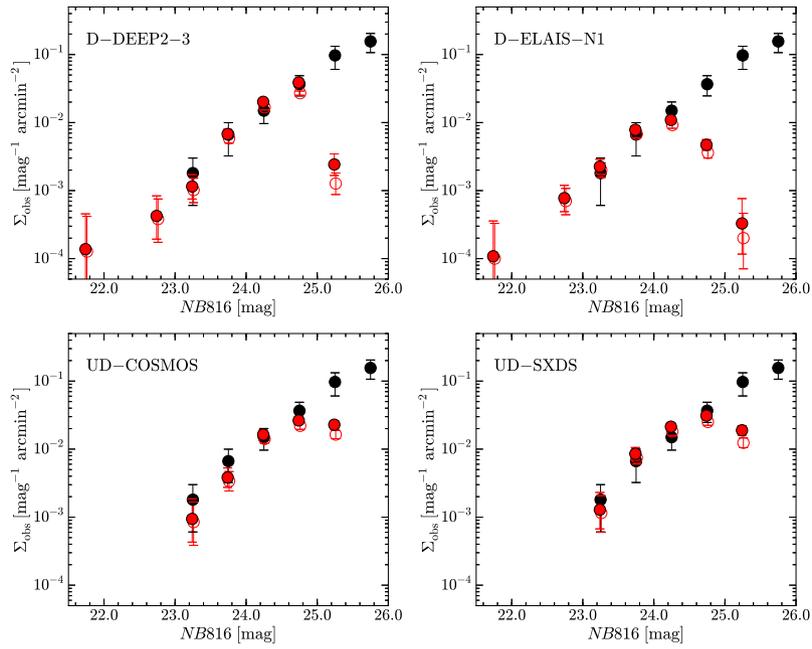}
\end{center}
\caption{
Surface number densities of $z = 5.7$ LAEs.
The red filled and open circles represent our surface number densities at $z = 5.7$ in each field 
with and without detection completeness correction (Section \ref{sec:LF_detcomp}), respectively.
The black circles denote the $z = 5.7$ surface number densities of 
\citet{2008ApJS..176..301O}. 
}\label{fig:LF_snd_z5p7}
\end{figure*}

\begin{figure*}
\begin{center}
\includegraphics[width=17cm]{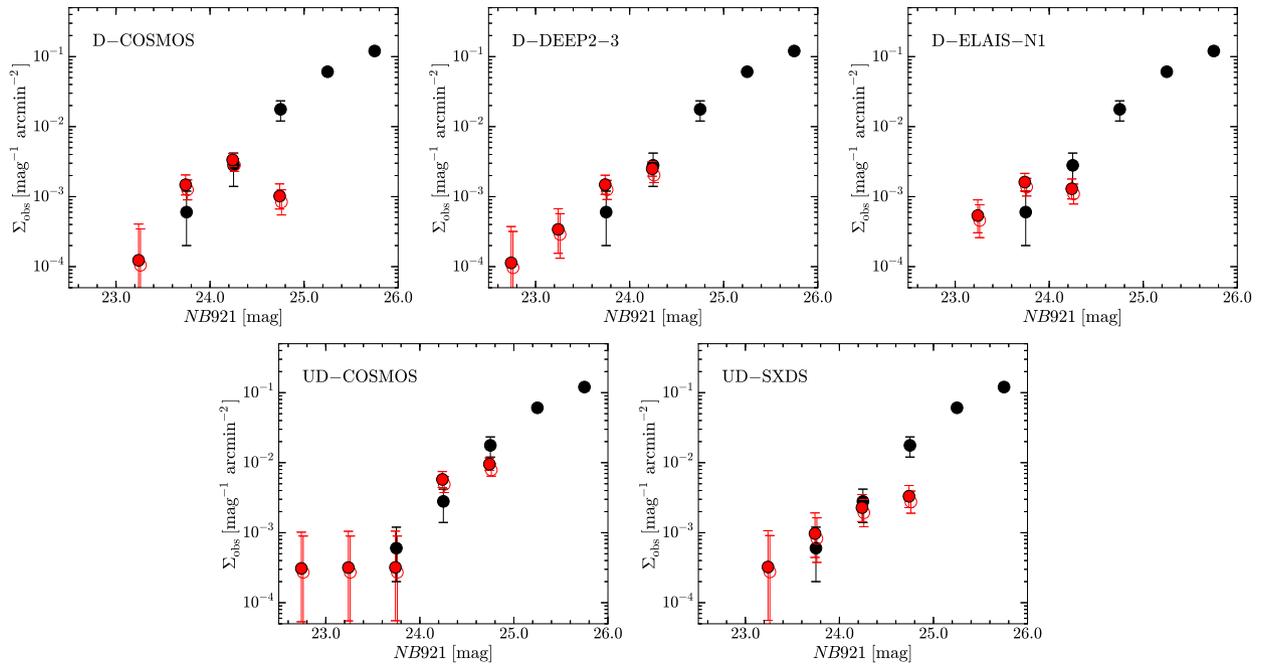}
\end{center}
\caption{
Same as Figure \ref{fig:LF_snd_z5p7}, but for $z = 6.6$. 
The black circles denote the $z = 6.6$ surface number densities of 
\citet{2010ApJ...723..869O}. 
}\label{fig:LF_snd_z6p6}
\end{figure*}

\subsection{Contamination}
\label{sec:LF_contami}

In our companion paper \citet{2017arXiv170500733S}, 
we estimate the contamination fractions in our $z = 5.7$ and $6.6$ LAE samples 
based on $81$ LAE candidates whose spectroscopic redshifts 
are obtained by our past and present programs 
with Subaru/Faint Object Camera and Spectrograph (FOCAS; \cite{2002PASJ...54..819K}), 
Magellan/Low Dispersion Survey Spectrograph 3 (LDSS3), 
and Magellan/Inamori Magellan Areal Camera and Spectrograph (IMACS; \cite{2011PASP..123..288D}). 
We find that 
$28$ ($53$) LAE candidates at $z=6.6$ ($z=5.7$) have been spectroscopically 
observed and $4$ out of the $28$ ($4$ out of the $53$) LAE candidates are 
found to be low-$z$ interlopers. 
Based on these results, 
the contamination fraction, $f_\mathrm{cont}$, is estimated to be 
$f_\mathrm{cont} = 4/28 \simeq 14${\%} 
($4/53 \simeq 8${\%}) for the $z=6.6$ ($z=5.7$) LAE sample. 
We also estimate the contamination fractions for bright LAE candidates 
with \textit{NB} $< 24$ mag. 
We have spectroscopically observed $18$ bright LAE candidates.  
Out of the $18$ candidates, $13$ sources are confirmed as LAEs 
and the other $5$ objects are strong [{\sc Oiii}] emitters at low $z$. 
Based on our spectroscopy results, the contamination rates 
for the bright $z=6.6$ and $z=5.7$ LAE samples 
are $f_\mathrm{cont} \simeq 33${\%} ($=4/12$) and $\simeq 17${\%} ($=1/6$), respectively. 
Although the contamination rates appear to depend on \textit{NB} magnitude, 
the estimated values are in the range of around $0-30${\%} 
and have large uncertainties due to the small 
number of our spectroscopically confirmed sources at this 
early stage of our program. In this study, 
we take into account this systematic uncertainty 
by increasing the lower $1\sigma$ confidence intervals of the Ly$\alpha$ LFs by $30${\%} 
(see Section \ref{sec:LF_z5p76p6}). 
Note that our estimated $f_\mathrm{cont}$ values are 
similar to those obtained in 
\citet{2008ApJS..176..301O}, \citet{2010ApJ...723..869O}, and \citet{2011ApJ...734..119K} 
($f_\mathrm{cont} = 0 - 30$\%),
who have conducted the Subaru/Suprime-Cam imaging survey for LAEs
at $z = 5.7$ and $6.6$.

\subsection{Surface Number Densities and Color Distributions}
\label{sec:LF_snd_cd}

\begin{figure*}
\begin{center}
\includegraphics[width=11cm]{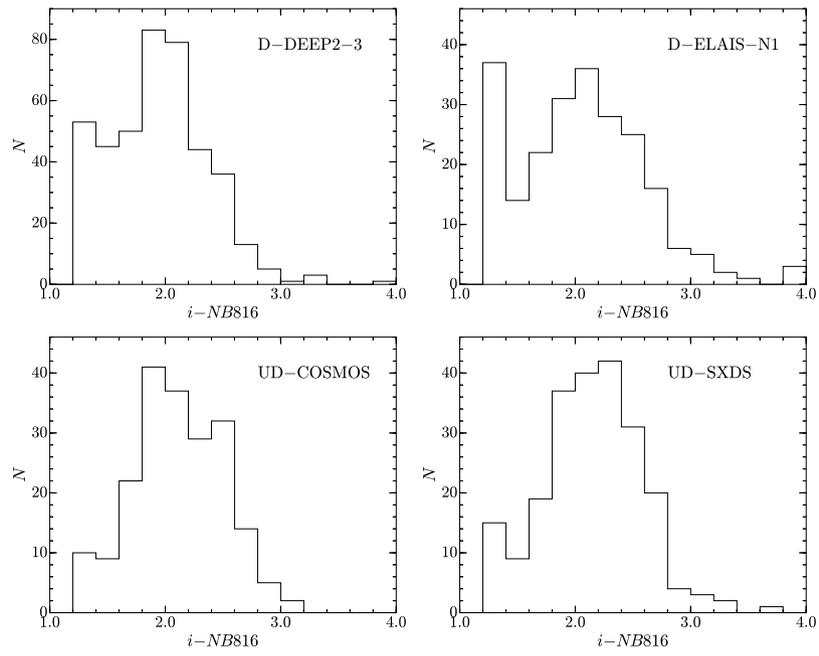}
\end{center}
\caption{
$i -$\textit{NB816} color distribution of $z = 5.7$ LAEs in each field. 
}\label{fig:LF_cd_z5p7}
\end{figure*}

\begin{figure*}
\begin{center}
\includegraphics[width=17cm]{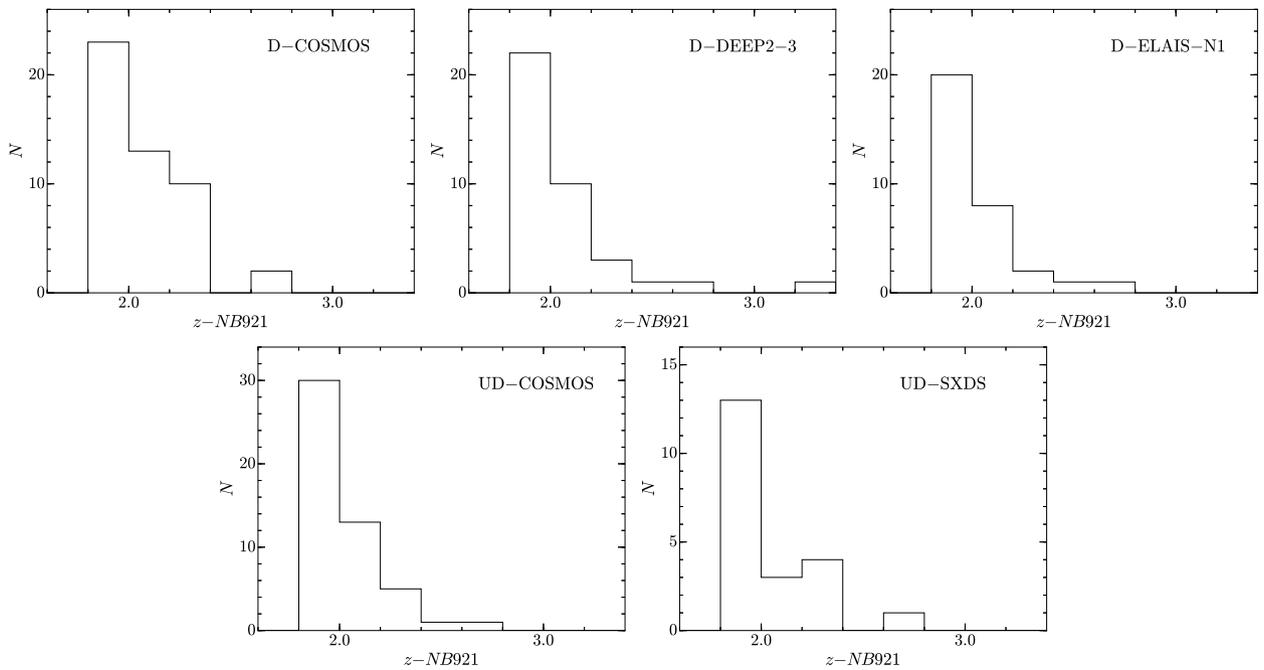}
\end{center}
\caption{
$z -$\textit{NB921} color distribution of 
$z = 6.6$ LAEs in each field. 
}\label{fig:LF_cd_z6p6}
\end{figure*}

Figures \ref{fig:LF_snd_z5p7} and \ref{fig:LF_snd_z6p6} represent the LAE surface number densities
at $z = 5.7$ and $6.6$, respectively, derived with our HSC SSP survey data.
We obtain the surface number densities by dividing the number counts of LAEs
by our survey areas (Section \ref{sec:sample_image}).
These surface number densities are corrected for the detection completeness (Section \ref{sec:LF_detcomp}). 
The $1\sigma$ error bars of the surface number densities are calculated 
based on the Poisson statistics \citep{1986ApJ...303..336G}, 
because the number counts of LAEs are small in some bright-end bins 
and their errors are not well represented by the square root values of the number counts. 
We use the Poisson single-sided limit values in the columns of ``0.8413'' in Tables 1 and 2 of 
\citet{1986ApJ...303..336G} 
for the $1\sigma$ upper and lower confidence intervals, respectively.
Note that the surface number densities decrease at faint magnitude bins 
due to the color-selection incompleteness.
For comparison,
we show the surface densities at $z = 5.7$ and $6.6$ of 
\citet{2008ApJS..176..301O} and \citet{2010ApJ...723..869O} 
in Figures \ref{fig:LF_snd_z5p7} and \ref{fig:LF_snd_z6p6}, respectively.
These previous studies have conducted deep narrowband imaging surveys for LAEs in the SXDS field, 
which is the sky region overlapping the UD-SXDS field in our HSC SSP survey.
In these figures, we find that our surface densities are broadly consistent with those of 
\citet{2008ApJS..176..301O} and \citet{2010ApJ...723..869O}.

Figures \ref{fig:LF_cd_z5p7} and \ref{fig:LF_cd_z6p6} show the color distributions of
$\textit{i} - \textit{NB816}$ and $z - \textit{NB921}$ for $z = 5.7$ and $6.6$ LAEs, respectively.
Magnitudes with a detection significance below $2 \sigma$ 
are replaced with the $2 \sigma$ limiting magnitudes. 
Based on Figures \ref{fig:LF_snd_z5p7}$-$\ref{fig:LF_cd_z6p6},
we estimate the best-fit Schechter functions and Ly$\alpha$ EW$_0$ distributions 
by the Monte Carlo simulations in Section \ref{sec:LF_compare}.

\subsection{Ly$\alpha$ Luminosity Functions at $z=5.7$ and $6.6$}
\label{sec:LF_z5p76p6}

We present Ly$\alpha$ LFs at $z = 5.7$ and $6.6$
based on our HSC Ly$\alpha$ LF samples 
constructed in Section \ref{sec:sample_sample}.
We derive the Ly$\alpha$ LFs in the same manner as 
\citet{2008ApJS..176..301O} and \citet{2010ApJ...723..869O}. 
We calculate the Ly$\alpha$ EW$_0$ values of $z = 5.7$ (6.6) LAEs 
from the magnitudes of \textit{NB816} (\textit{NB921}) and $z$ band,
and estimate the Ly$\alpha$ luminosities of LAEs 
from these EW$_0$ values and the total magnitudes of \textit{NB816} (\textit{NB921}), 
under the assumption that
the spectrum of LAEs has a Ly$\alpha$ line and 
a flat UV continuum (i.e., $f_{\nu} =$ constant) 
with the IGM absorption of \citet{1995ApJ...441...18M}, 
following the methods described in 
\citet{2006PASJ...58..313S}, \citet{2010ApJ...723..869O}, and \citet{2014ApJ...797...16K}. 
Ly$\alpha$ luminosities are calculated, 
assuming that Ly$\alpha$ emission is placed at the central wavelength of the narrowbands.
The uncertainties of the Ly$\alpha$ luminosities are calculated 
based on the uncertainties of the \textit{NB} and $z$ band magnitudes. 
We obtain the volume number density of LAEs in each Ly$\alpha$ luminosity bin 
by dividing the number of observed LAEs in each bin by our survey volume 
(Section \ref{sec:sample_image}).
We correct these number densities for the detection 
completeness estimated in Section \ref{sec:LF_detcomp}.
The $1\sigma$ uncertainties of the Lya LF measurements are calculated based on Poisson statistics 
\citep{1986ApJ...303..336G}.  
Note that we do not include the field-to-field variance in the uncertainties of our Ly$\alpha$ LFs,
because the survey areas for $z = 5.7$ and $6.6$ LAEs are very large
(see Section \ref{sec:sample_image}).
This procedure of Ly$\alpha$ LF derivation is known as the classical method.

\begin{figure*}
\begin{center}
\includegraphics[width=17cm]{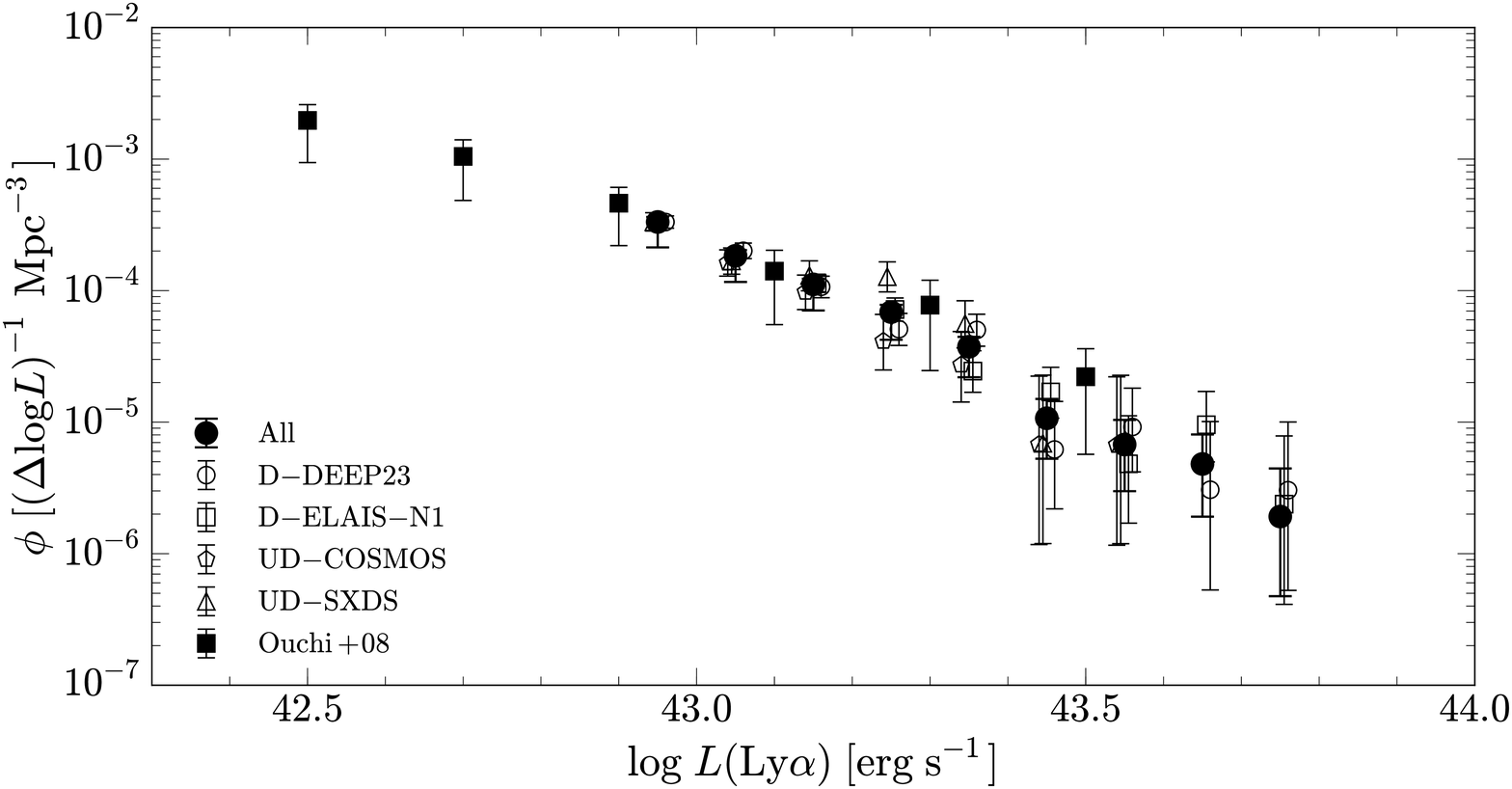}
\includegraphics[width=17cm]{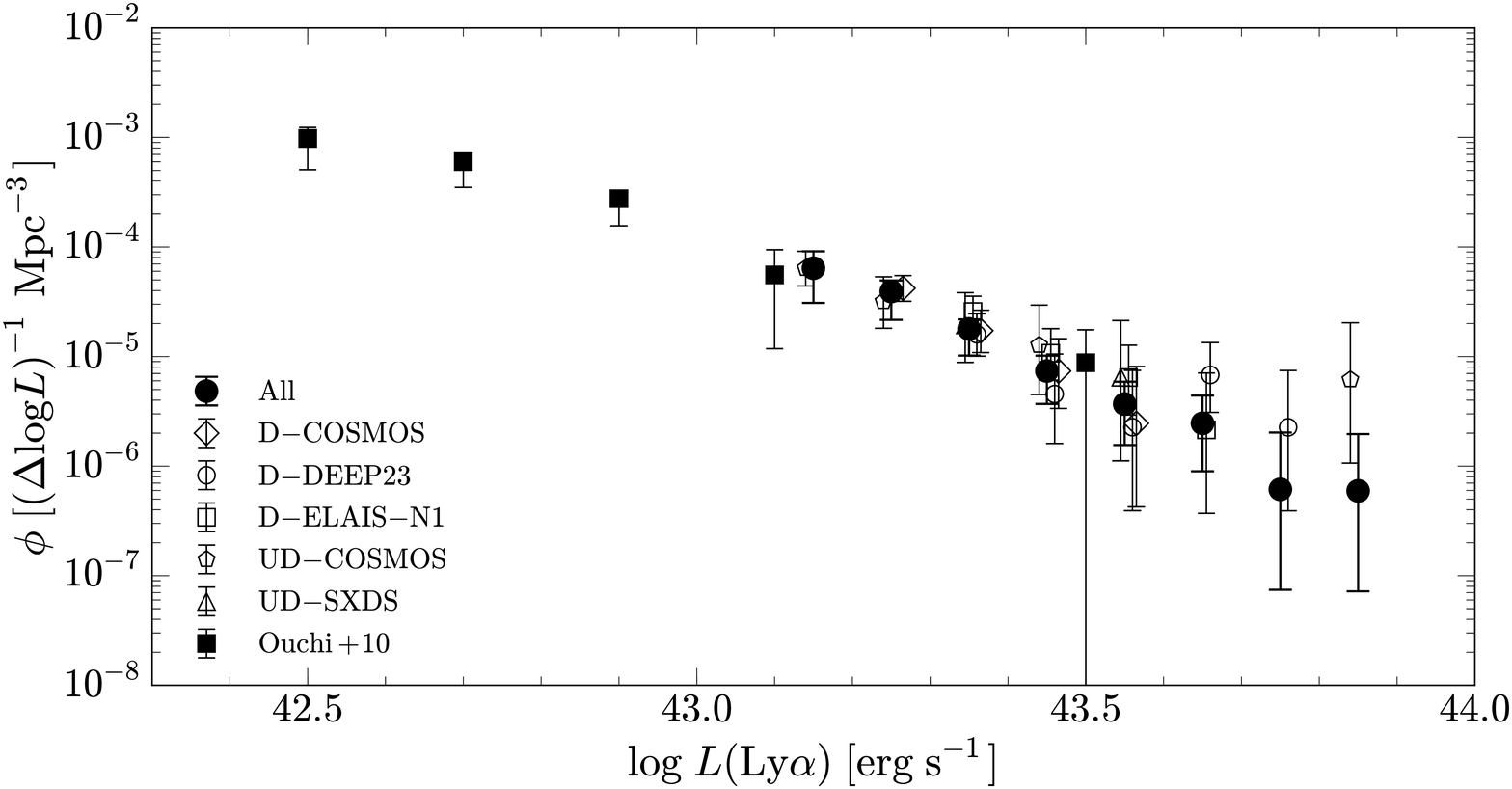}
\end{center}
\caption{
Ly$\alpha$ LFs of LAEs at $z = 5.7$ (top) and $z = 6.6$ (bottom).
The filled circles represent our results averaged over the separate fields. 
The open circles, squares, pentagons, triangles, and diamonds are 
the Ly$\alpha$ LFs of the separate fields of 
D-DEEP2-3, D-ELAIS-N1, UD-COSMOS, UD-SXDS, and D-COSMOS, respectively. 
The filled squares are the results of \citet{2008ApJS..176..301O} and \citet{2010ApJ...723..869O}. 
}\label{fig:LF_each_field}
\end{figure*}

We first show 
our derived Ly$\alpha$ LFs at $z = 5.7$ and $z = 6.6$ 
with the classical method in Figure \ref{fig:LF_each_field}. 
To check field-to-field variations, 
we present the $z = 5.7$ and $z = 6.6$ Ly$\alpha$ LF results for the four and five fields 
in the top and bottom panels, respectively, 
as well as the results averaged over these fields. 
We find that our results for these separate fields 
are consistent with each other, although they have 
relatively large uncertainties.

\begin{figure*}
\begin{center}
\includegraphics[width=17cm]{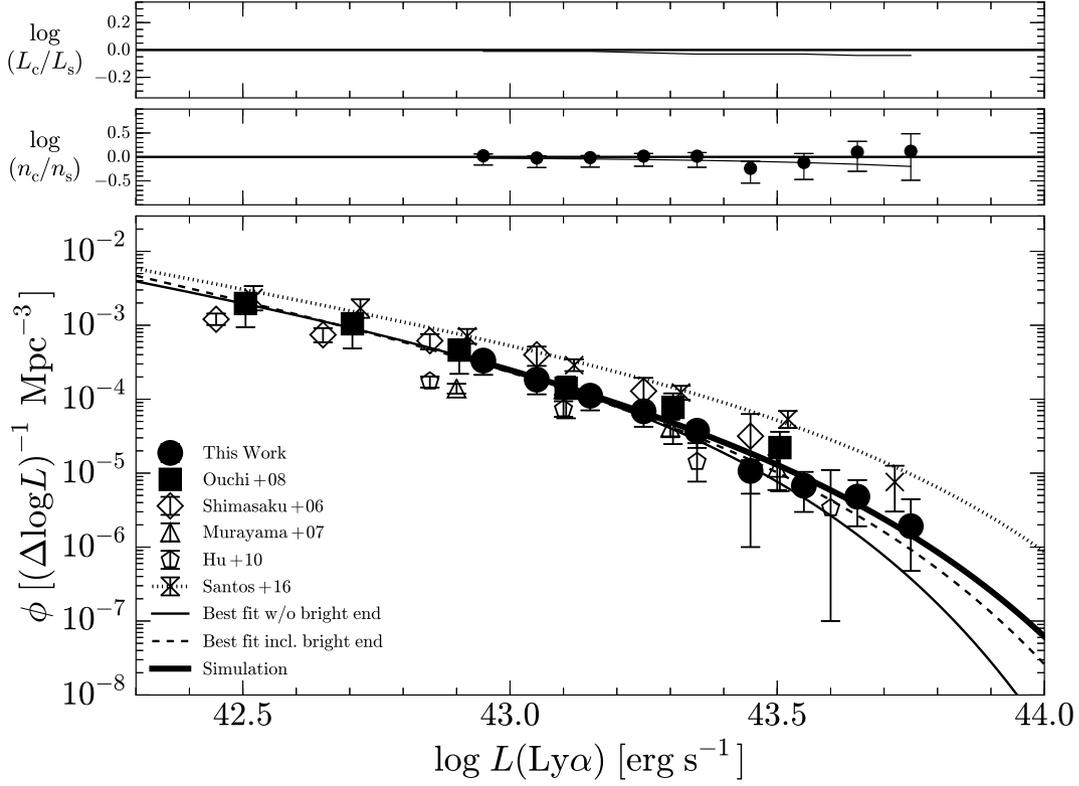}
\end{center}
\caption{
\textit{Top}:  
Ratio of the luminosity from the classical method ($L_\mathrm{c}$)
to that from the simulations ($L_\mathrm{s}$) 
at the same number density as a function of $L_\mathrm{c}$ 
based on comparisons of the best-fit LFs.  
\textit{Middle}: 
Ratio of the number density from the classical method ($n_\mathrm{c}$)
to that from the simulations ($n_\mathrm{s}$) at a given luminosity. 
The filled circles compare 
the LF data points derived by the classical method (filled circles in the bottom panel)
with the best-fit LF derived by the simulations (thick curve in the bottom panel). 
The solid curve is based on a comparison of 
the two best-fit LFs obtained by the classical method and the simulations. 
\textit{Bottom}: 
Ly$\alpha$ LFs of $z = 5.7$ LAEs.
The filled circles represent our $z = 5.7$ Ly$\alpha$ LF results based on the HSC SSP data, 
where we consider the contamination fraction $f_\mathrm{cont} = 0-30$\% in the LF uncertainties.  
The filled squares denote the Ly$\alpha$ LF given by 
\citet{2008ApJS..176..301O}. 
The best-fit Schechter function for the Ly$\alpha$ LFs of our and 
\authorcite{2008ApJS..176..301O}'s 
studies are shown with the thin (dashed) curve, 
where the Ly$\alpha$ luminosity range of $\log L(\mathrm{Ly}\alpha)$ [erg s$^{-1}$] $= 42.4 - 43.5$ ($44.0$) is considered. 
We also show the end-to-end Monte Carlo simulation result, as described in Section \ref{sec:LF_z5p76p6}, with the thick curve.
The open diamonds, triangles, pentagons, and crosses are 
the Ly$\alpha$ LFs of 
\citet{2006PASJ...58..313S}, \citet{2007ApJS..172..523M}, 
\citet{2010ApJ...725..394H}, and \citet{2016MNRAS.463.1678S}, respectively.
}\label{fig:LF_thisstudy_compare_z5p7}
\end{figure*}

\begin{figure*}
\begin{center}
\includegraphics[width=17cm]{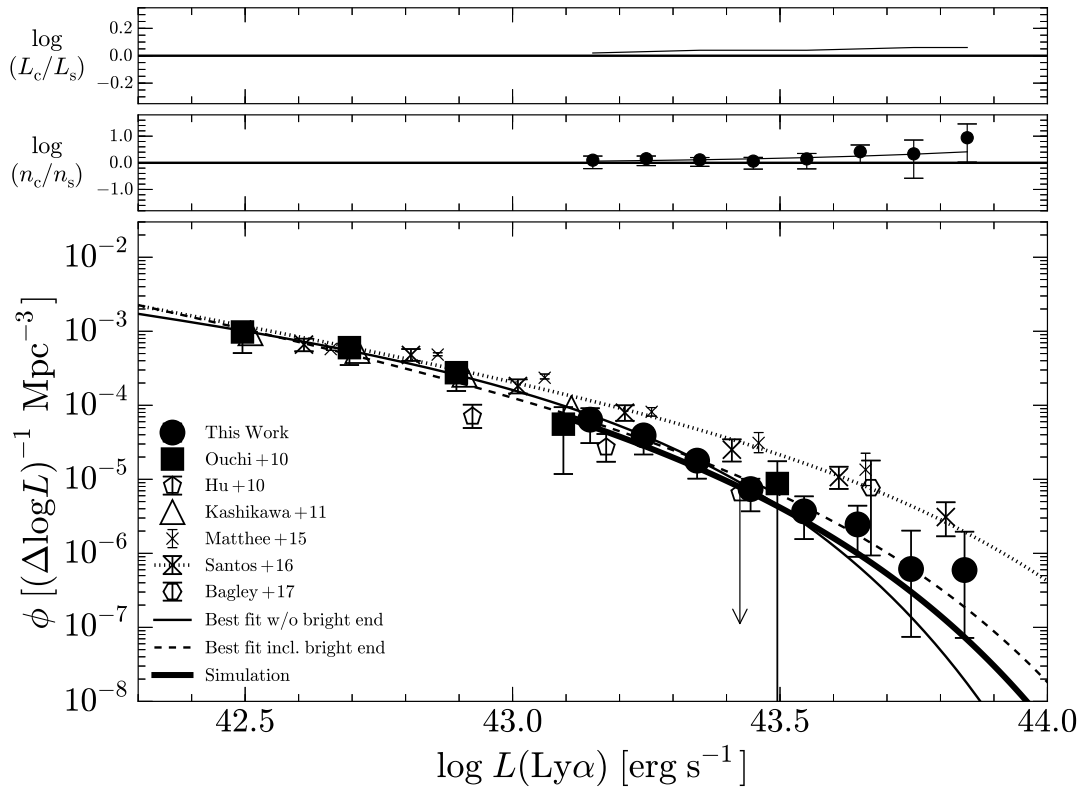}
\end{center}
\caption{
Same as Figure \ref{fig:LF_thisstudy_compare_z5p7}, but for $z = 6.6$.
In the bottom panel, 
the filled squares denote the Ly$\alpha$ LF given by \citet{2010ApJ...723..869O}. 
The thin (dashed) curve is 
the best-fit Schechter function for the Ly$\alpha$ LFs of our and \authorcite{2010ApJ...723..869O}'s results, 
where the Ly$\alpha$ luminosity range of $\log L(\mathrm{Ly}\alpha)$ [erg s$^{-1}$] $= 42.4 - 43.5$ ($44.0$) is taken into account.  
The simulation result is also shown with the thick curve.
The open pentagons and crosses are the Ly$\alpha$ LFs of \citet{2010ApJ...725..394H}.
The small crosses are the Ly$\alpha$ LFs of \citet{2015MNRAS.451..400M} 
derived for the UDS and COSMOS fields. 
The large crosses are taken from \citet{2016MNRAS.463.1678S}, 
who have revised the Ly$\alpha$ LFs of \citet{2015MNRAS.451..400M}. 
The open triangles are the Ly$\alpha$ LF obtained by \citet{2011ApJ...734..119K}. 
The open pentagon represents the Ly$\alpha$ LF at $z = 6.4$ by \citet{2017ApJ...837...11B}, 
who have conducted \textit{HST} WFC3 infrared spectroscopic parallel survey.
}\label{fig:LF_thisstudy_compare_z6p6}
\end{figure*}

In Figure \ref{fig:LF_thisstudy_compare_z5p7}, 
we show 
our Ly$\alpha$ LF at $z = 5.7$ derived with the classical method 
and previous results.
The filled circles 
represent our $z = 5.7$ Ly$\alpha$ LF, which is derived from the HSC SSP data.
Our Ly$\alpha$ LF covers a Ly$\alpha$ luminosity range of
$\log L(\mathrm{Ly}\alpha)$ [erg s$^{-1}$] $= 42.9 - 43.8$.
The wide area of the HSC SSP survey 
allows us to probe this brighter luminosity range than those of previous studies 
(e.g., \cite{2006PASJ...58..313S}; \cite{2008ApJS..176..301O}; \cite{2010ApJ...725..394H}). 
We take into account 
the contamination fractions in our samples (Section \ref{sec:LF_contami}) 
in the calculations of the Ly$\alpha$ LF uncertainties 
by increasing the lower $1\sigma$ confidence intervals by $30${\%}. 
Similarly, 
in Figure \ref{fig:LF_thisstudy_compare_z6p6}, 
we show our $z = 6.6$ Ly$\alpha$ LF from the HSC SSP data derived with the classical method. 
The uncertainties from the $f_\mathrm{cont}$ value (Section \ref{sec:LF_contami}) are considered. 
Our $z = 6.6$ Ly$\alpha$ LF covers a bright Ly$\alpha$ luminosity range of
$\log L(\mathrm{Ly}\alpha)$ [erg s$^{-1}$] $= 43.0 - 43.8$
thanks to the wide area of the HSC SSP survey. 
Table \ref{table:LyaLF_data} shows the values of 
our Ly$\alpha$ LFs at $z=5.7$ and $z=6.6$.

We fit a Schechter function 
\citep{1976ApJ...203..297S}
to our $z = 5.7$ 
and $z=6.6$ 
Ly$\alpha$ LFs
by minimum $\chi^2$ fitting.
The Schechter function is defined by
\begin{equation}
\phi (L_{\mathrm{Ly}\alpha}) dL_{\mathrm{Ly}\alpha} =  \phi^{*}_{\mathrm{Ly}\alpha} \left( \frac{L_{\mathrm{Ly}\alpha}}{L^{*}_{\mathrm{Ly}\alpha}} \right)^{\alpha} \exp \left( - \frac{L_{\mathrm{Ly}\alpha}}{L^{*}_{\mathrm{Ly}\alpha}} \right) d \left( \frac{L_{\mathrm{Ly}\alpha}}{L^{*}_{\mathrm{Ly}\alpha}} \right),	\label{eq:schechter}
\end{equation}
where $L^{*}_{\mathrm{Ly}\alpha}$ is the characteristic Ly$\alpha$ luminosity, 
$\phi^{*}_{\mathrm{Ly}\alpha}$ is the normalization, 
and $\alpha$ is the faint-end slope. 
We consider two cases. In one case, 
we use our Ly$\alpha$ LF measurements 
at $\log L(\mathrm{Ly}\alpha)$ [erg s$^{-1}$] $< 43.5$, 
where AGN contamination is not significant in lower-$z$ LAE studies \citep{2008ApJS..176..301O,2016ApJ...823...20K}.\footnote{
As mentioned in Section \ref{sec:discuss_bright}, 
\citet{2017arXiv170500733S} have found no clear 
signature of AGNs 
for several bright LAEs with $\log L(\mathrm{Ly}\alpha)$ [erg s$^{-1}$] $> 43.5$. 
Their bright LAEs show narrow Ly$\alpha$ line widths of $< 400$ km s$^{-1}$ 
and no clear detection of UV lines such as {\sc N v} and {\sc C iv}. 
However, their investigation is based on the rest-frame UV spectroscopic observations 
and they cannot rule out the possibility that 
the bright LAEs host an AGN with faint highly-ionized UV lines 
(e.g., \cite{2004AJ....127.3146H}; \cite{2006MNRAS.370.1479M}). 
In this paper, 
we present the Ly$\alpha$ LF fitting results 
for the two cases where we include and exclude 
the bright-end bins of 
$\log L(\mathrm{Ly}\alpha)$ [erg s$^{-1}$] $> 43.5$ 
for a conservative discussion. 
}
In the other case, 
we include the bright-end LF
results at $\log L(\mathrm{Ly}\alpha)$ [erg s$^{-1}$] $\geq 43.5$. 
In both of these cases, 
we also use the faint-end Ly$\alpha$ LFs of 
\citet{2008ApJS..176..301O} and \citet{2010ApJ...723..869O} 
for $z=5.7$ and $z=6.6$, respectively. 
This is because the faint-end Ly$\alpha$ LFs of these studies 
cover faint Ly$\alpha$ luminosity ranges that we do not reach. 
Specifically, we include 
the $z=5.7$ Ly$\alpha$ LF data points of \citet{2008ApJS..176..301O} 
in the range of $\log L(\mathrm{Ly}\alpha)$ [erg s$^{-1}$] $= 42.4 - 42.9$ 
and the $z=6.6$ Ly$\alpha$ LF data points of \citet{2010ApJ...723..869O} 
in the range of $\log L(\mathrm{Ly}\alpha)$ [erg s$^{-1}$] $= 42.4 - 43.0$, 
both of which are not overlapped with the luminosity ranges of our derived LFs. 
The best-fit Schechter function parameters are listed in Table \ref{table:LF_schechter} 
and 
the best-fit Schechter functions are shown in 
Figures \ref{fig:LF_thisstudy_compare_z5p7} and \ref{fig:LF_thisstudy_compare_z6p6} 
(black thin curve and dashed curve).

\begin{table}
\caption{
Ly$\alpha$ LFs at $z = 5.7$ and $6.6$ from this work 
}
\begin{center}
\begin{threeparttable}
\small
\begin{tabular}{cc}
\hline
\hline
$\log L(\mathrm{Ly}\alpha)$\tnote{a}		& $\log \phi$\tnote{b}	\\
(erg s$^{-1}$) 						& ([$\Delta \log L(\mathrm{Ly}\alpha)$]$^{-1}$ Mpc$^{-3}$)	\\
\hline
\multicolumn{2}{c}{$z = 5.7$}	\\
\hline
$42.95$	& ${-3.478}^{+0.038}_{-0.193}$	\\
$43.05$	& ${-3.735}^{+0.044}_{-0.199}$	\\
$43.15$	& ${-3.953}^{+0.043}_{-0.198}$	\\
$43.25$	& ${-4.163}^{+0.055}_{-0.210}$	\\
$43.35$	& ${-4.427}^{+0.076}_{-0.231}$	\\
$43.45$	& ${-4.970}^{+0.147}_{-0.308}$	\\
$43.55$	& ${-5.170}^{+0.187}_{-0.355}$	\\
$43.65$	& ${-5.318}^{+0.224}_{-0.401}$	\\
$43.75$	& ${-5.717}^{+0.365}_{-0.606}$	\\
\hline
\multicolumn{2}{c}{$z = 6.6$}	\\
\hline
$43.15$	& ${-4.194}^{+0.154}_{-0.317}$	\\
$43.25$	& ${-4.407}^{+0.101}_{-0.258}$	\\
$43.35$	& ${-4.748}^{+0.087}_{-0.243}$	\\
$43.45$	& ${-5.132}^{+0.140}_{-0.300}$	\\
$43.55$	& ${-5.433}^{+0.203}_{-0.374}$	\\
$43.65$	& ${-5.609}^{+0.253}_{-0.438}$	\\
$43.75$	& ${-6.212}^{+0.519}_{-0.917}$	\\
$43.85$	& ${-6.226}^{+0.519}_{-0.917}$	\\
\hline
\end{tabular}\label{table:LyaLF_data}
\begin{tabnote}
\item[a] The luminosity bin of our Ly$\alpha$ LFs at $z = 5.7$ and $6.6$.
The bin size is $\Delta \log L(\mathrm{Ly}\alpha) = 0.1$.
\item[b] The number densities corrected for the detection completeness
(see Section \ref{sec:LF_detcomp}).
\end{tabnote}
\end{threeparttable}
\end{center}
\end{table}

\begin{table*}
\caption{
Best-fit Schechter Parameters and Ly$\alpha$ Luminosity Densities}
\begin{center}
\begin{threeparttable}
\small
\begin{tabular}{lcccc}
\hline
\hline
Redshift	& $L^{*}$		 			& $\phi^{*}$				& $\alpha$ 				& $\rho^{{\mathrm{Ly}\alpha}}_{\rm obs}$\tnote{a}	\\
		& ($10^{43}$ erg s$^{-1}$)	& ($10^{-4}$ Mpc$^{-3}$)		& 						& ($10^{39}$ erg s$^{-1}$ Mpc$^{-3}$)	\\
\hline
\multicolumn{5}{c}{Classical method for $\log L(\mathrm{Ly}\alpha)$ [erg s$^{-1}$] $= 42.4 - 43.5$\tnote{b}}	\\
\hline
5.7		& $1.07^{+0.77}_{-0.38}$		& $2.46^{+3.48}_{-1.86}$		& $-2.26^{+0.76}_{-0.44}$ 	& $3.39$			\\
6.6		& $0.82^{+0.86}_{-0.30}$		& $2.83^{+3.52}_{-2.38}$		& $-1.86^{+0.79}_{-0.67}$ 	& $1.96$			\\
\hline
\multicolumn{5}{c}{Classical method for $\log L(\mathrm{Ly}\alpha)$ [erg s$^{-1}$] $= 42.4 - 44.0$\tnote{c}}	\\
\hline
5.7		& $1.64^{+2.16}_{-0.62}$		& $0.849^{+1.87}_{-0.771}$	& $-2.56^{+0.53}_{-0.45}$ 	& $3.49$			\\
6.6		& $1.66^{+0.30}_{-0.69}$		& $0.467^{+1.44}_{-0.442}$	& $-2.49^{+0.50}_{-0.50}$ 	& $1.82$			\\
\hline
\multicolumn{5}{c}{End-to-end Monte Carlo simulations}	\\
\hline
5.7		& $2.0$		& $0.63$		& $-2.6$ (fix) 	& $3.5$			\\
6.6		& $1.3$		& $0.63$		& $-2.5$ (fix) 	& $1.7$			\\
\hline
\end{tabular}\label{table:LF_schechter}
\begin{tabnote}
\item[a] The Ly$\alpha$ luminosity densities obtained by integrating the Ly$\alpha$ LF
down to $\log L_{\mathrm{Ly}\alpha}$ [erg s$^{-1}$] $= 42.4$,
which corresponds to $\sim 0.3 \times L^{*}_{\mathrm{Ly}\alpha} (z=3-6)$. 
\item[b] The best-fit parameters are derived with the classical method 
for the Ly$\alpha$ LF measurements in the luminosity range of $\log L(\mathrm{Ly}\alpha)$ [erg s$^{-1}$] $= 42.4-43.5$. 
\item[c] The best-fit parameters are derived with the classical method 
for the Ly$\alpha$ LF measurements in the luminosity range of $\log L(\mathrm{Ly}\alpha)$ [erg s$^{-1}$] $= 42.4-44.0$. 
\end{tabnote}
\end{threeparttable}
\end{center}
\end{table*}

\begin{figure*}
\begin{center}
\includegraphics[width=8cm]{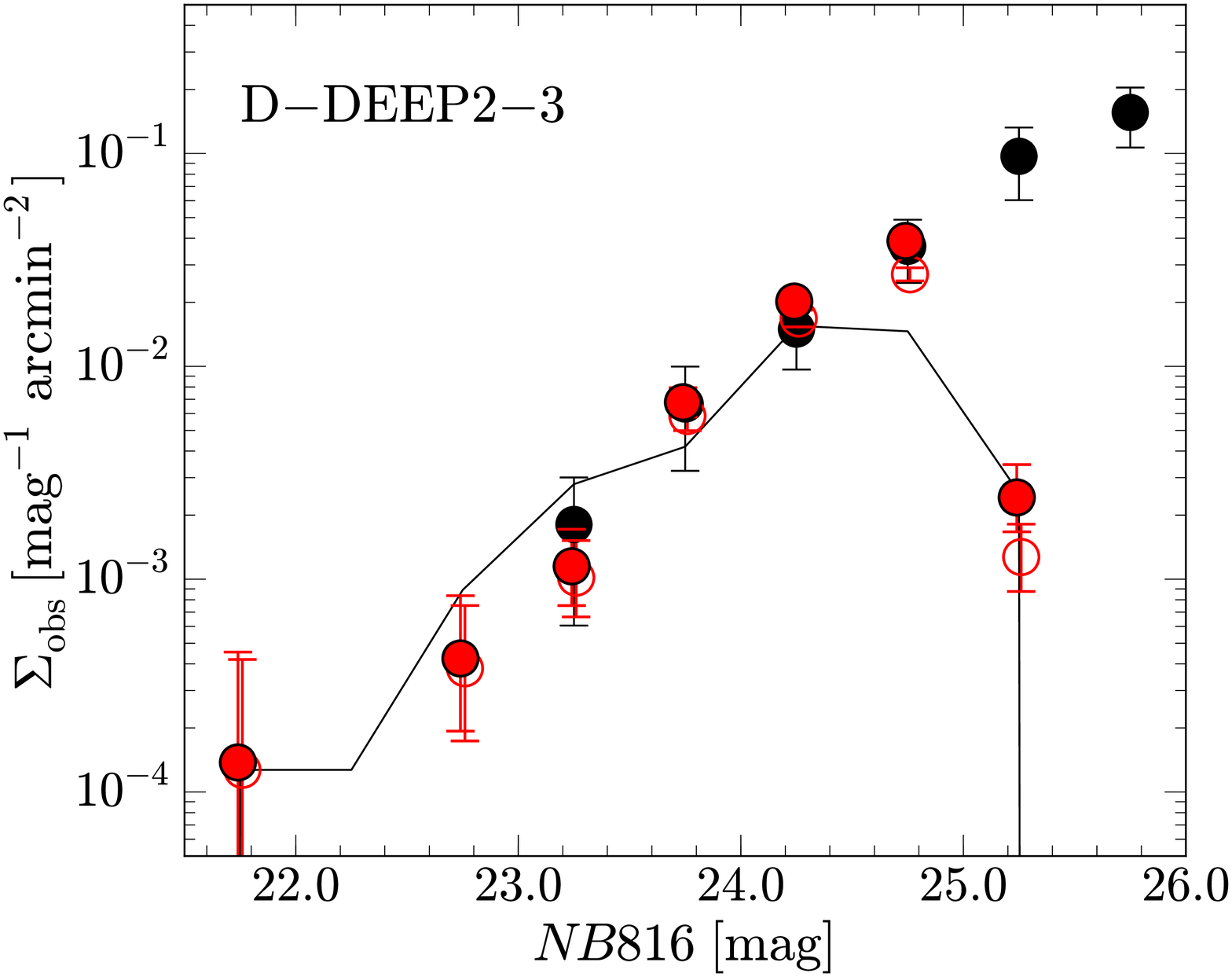}
\includegraphics[width=8cm]{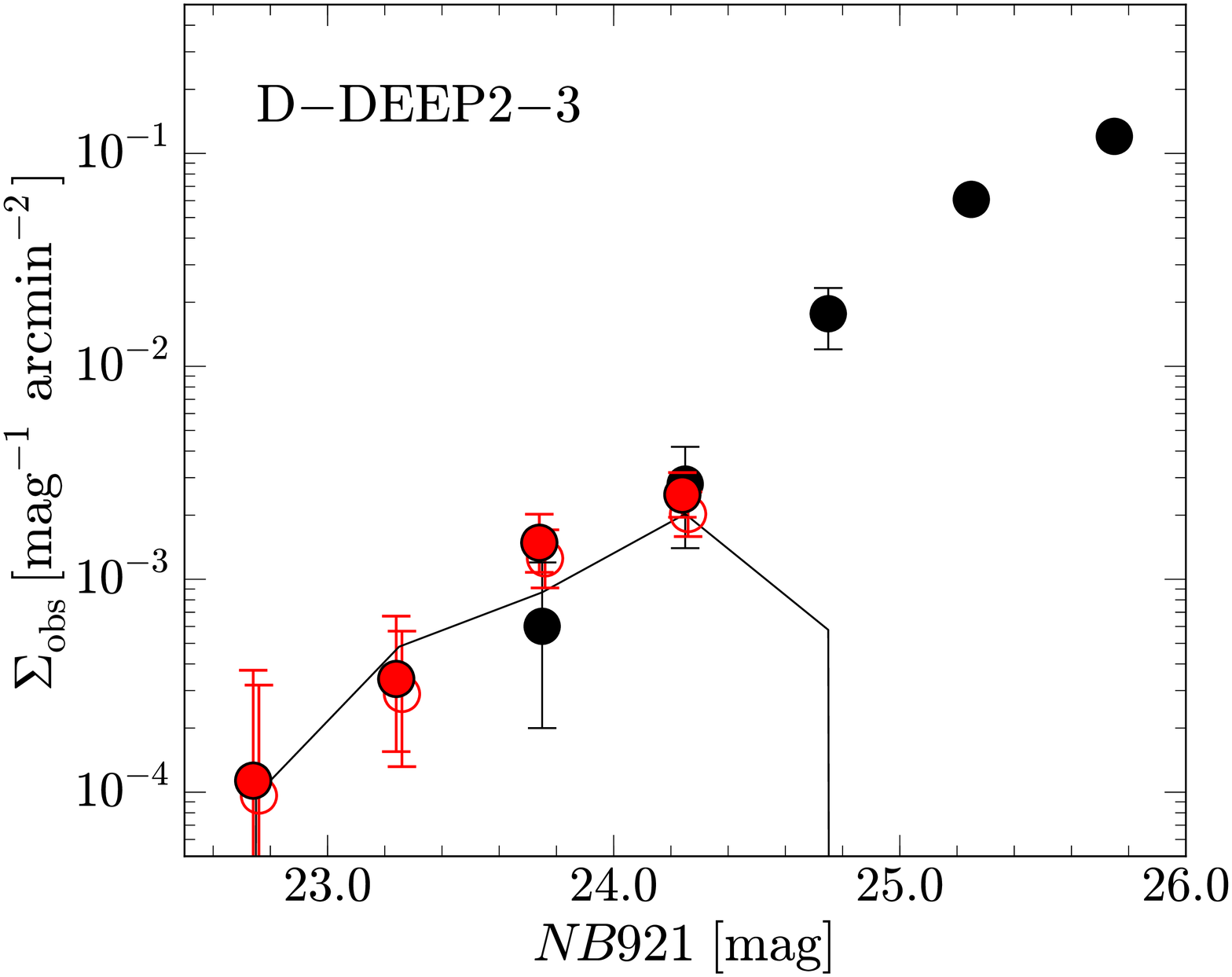}
\includegraphics[width=8cm]{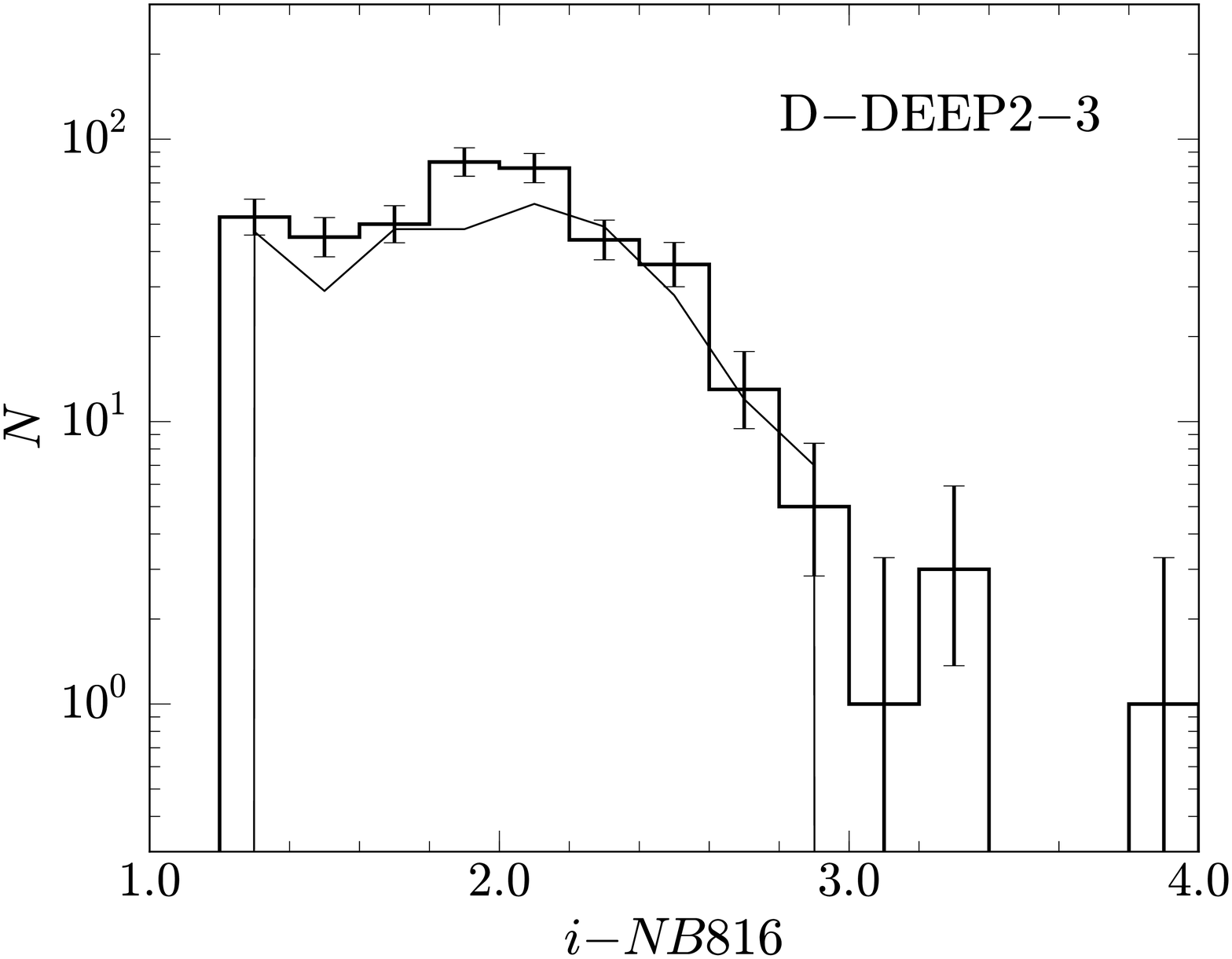}
\includegraphics[width=8cm]{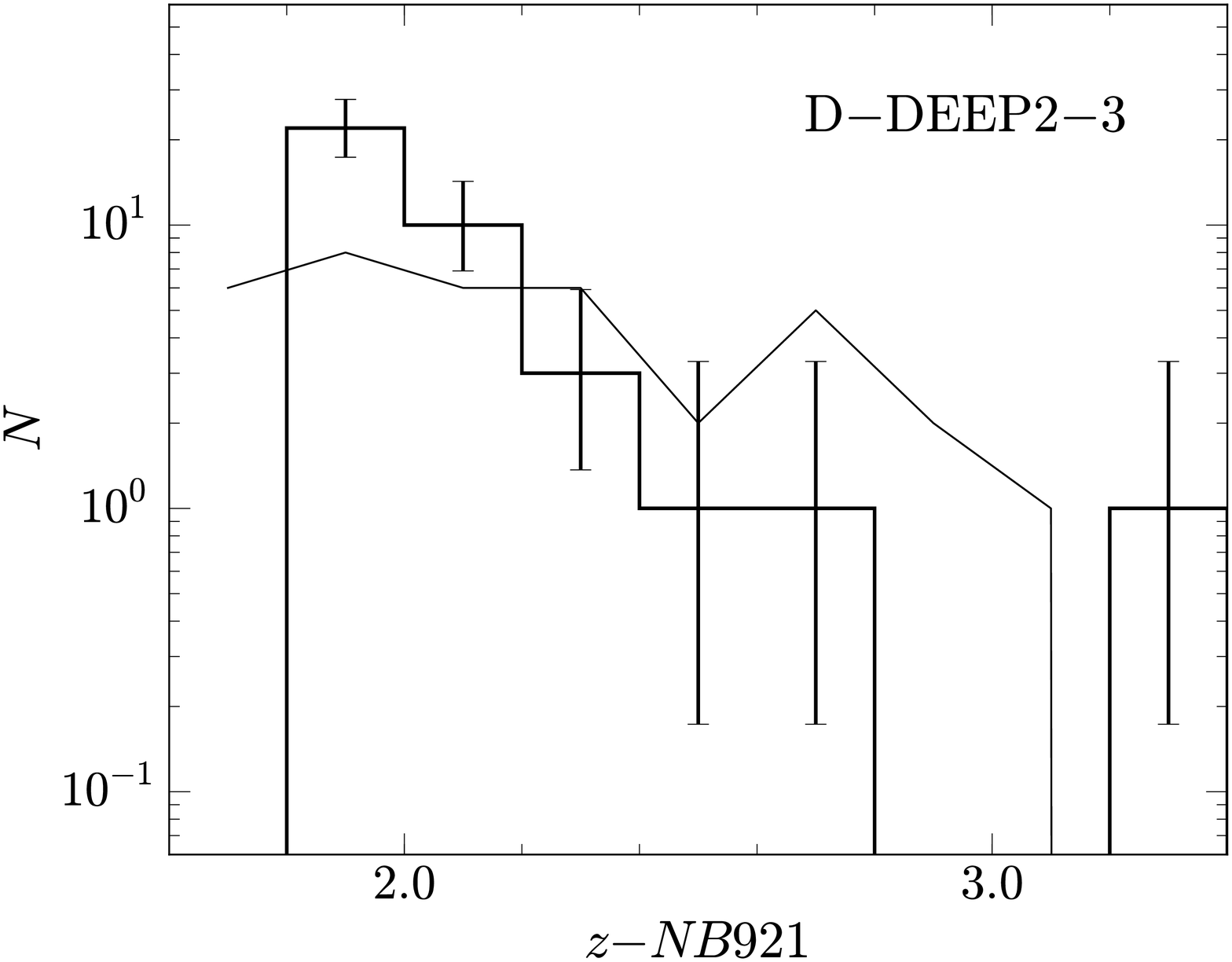}
\end{center}
\caption{
Examples of the results of our end-to-end Monte Carlo simulations. 
Top: best-fit surface number densities 
for the D-DEEP2-3 LAEs at $z=5.7$ (left) and $z=6.6$ (right) 
are shown with solid lines. 
The other symbols are the same as in Figures \ref{fig:LF_snd_z5p7} and \ref{fig:LF_snd_z6p6}. 
Bottom: best-fit \textit{BB}$-$\textit{NB} color distributions of LAEs. 
In the left and right panels, 
the solid lines denote the results for 
our $z=5.7$ and $z=6.6$ LAEs in the D-DEEP2-3 field, respectively. 
The other symbols are the same as in Figures \ref{fig:LF_cd_z5p7} and \ref{fig:LF_cd_z6p6}. 
}\label{fig:snd_cd_fit}
\end{figure*}

The classical method is accurate 
if the narrowband filter has an ideal boxcar transmission shape. 
However, the actual narrowband filter transmission shapes are close to a triangle, 
which causes mainly the following two systematic uncertainties in Ly$\alpha$ LF estimates by the classical method.  
(I) A Ly$\alpha$ flux of a LAE at a given narrowband magnitude depends on the redshift of the LAE.
(II) The minimum EW$_0$ value that corresponds to a given $\textit{BB}-\textit{NB}$ color criterion changes with redshift. 
These two systematic effects are closely related to each other. 
Moreover, there are many other systematic uncertainties 
including the survey volume definitions.
We evaluate such systematic uncertainties in our HSC Ly$\alpha$ LFs  
by carrying out end-to-end Monte Carlo simulations 
that are conducted in 
\citet{2006PASJ...58..313S} and \citet{2008ApJS..176..301O}.  
We generate a mock catalog of LAEs 
with a given set of Schechter function parameters ($\phi^{*}_{\mathrm{Ly}\alpha}$, $L^{*}_{\mathrm{Ly}\alpha}$, $\alpha$)
and a standard deviation ($\sigma$) of a Gaussian Ly$\alpha$ EW$_0$ probability distribution.
LAEs in the mock catalog are uniformly distributed 
in a comoving volume over the redshift range that a narrowband covers, 
and their narrowband and broadband magnitudes are measured. 
We then select LAEs using the same criteria as used for our LAE selections from the actual HSC data. 
Finally, we derive the surface number densities and color distributions 
of the selected LAEs,
and compare these results with the actual ones 
(see \cite{2006PASJ...58..313S} and \cite{2008ApJS..176..301O} 
for more details of the simulations). 
In this comparison, 
we use the surface number densities and color distributions 
that are obtained for the $z=5.7$ ($z=6.6$) LAEs 
in the four (five) fields separately  
to take into account the different relative depths of these fields. 
Free parameters in our end-to-end Monte Carlo simulations are 
$L^{*}_{\mathrm{Ly}\alpha}$ and $\phi^{*}_{\mathrm{Ly}\alpha}$ of the Schechter funtions  
and $\sigma$ of Gaussian Ly$\alpha$ EW$_0$ probability distributions. 
The faint-end slope $\alpha$ is fixed at $\alpha = -2.6$ for $z=5.7$ and $\alpha = -2.5$ for $z=6.6$, 
which are the same as those obtained with the classical method 
for the Ly$\alpha$ LF measurements in the range of 
$\log L(\mathrm{Ly}\alpha)$ [erg s$^{-1}$] $= 42.4-44.0$. 
Comparing the surface number densities (Figure \ref{fig:LF_snd_z5p7}) 
and color distributions (Figure \ref{fig:LF_cd_z5p7}) from the real data 
with those from the Monte Carlo simulations, 
we search for the best-fitting set of the three parameters that minimizes $\chi^2$. 
The best-fit Schechter parameters are summarized in Table \ref{table:LF_schechter} 
and examples of the fitting results are shown in Figure \ref{fig:snd_cd_fit}.

We show the best-fit functions from the Monte Carlo simulations 
for our Ly$\alpha$ LFs at $z = 5.7$ and $6.6$
in Figures \ref{fig:LF_thisstudy_compare_z5p7} and \ref{fig:LF_thisstudy_compare_z6p6}, respectively.
We find that the best-fit Schechter functions from the simulations
are consistent with our HSC Ly$\alpha$ LFs derived by the classical 
method.
Similar conclusions are obtained by 
\citet{2006PASJ...58..313S} and \citet{2008ApJS..176..301O}, 
who have derived the Ly$\alpha$ LFs at $z \sim 3-6$ with Subaru/Suprime-Cam.
We confirm that the classical method for the Ly$\alpha$ LF calculations  
gives a good approximation to the true Ly$\alpha$ LF even in the case of our HSC SSP data. 
The top panel of Figure \ref{fig:LF_thisstudy_compare_z5p7} 
compares the luminosities from the classical method ($L_{\rm c}$) and from the simulations ($L_{\rm s}$) 
at the same number densities as a function of $L_{\rm c}$. 
We find that 
the difference between these two luminosities is only $\lesssim$ 0.1 dex. 
Similarly, 
the middle panel of Figure \ref{fig:LF_thisstudy_compare_z5p7} 
shows the ratios of the number densities derived from the classical method 
to those from the simulations. 
We find that this ratio is also nearly 
equal to unity, where the departures of the classical-method data points 
from the simulation results are smaller than the statistical $\sim 1 \sigma$ 
uncertainties shown with the error bars.
Moreover, we also find that the classical-method data points 
are not always underestimated 
(Figures \ref{fig:LF_thisstudy_compare_z5p7}  vs. \ref{fig:LF_thisstudy_compare_z6p6} ). 
We thus think that the large correction factors 
beyond our statistical errors should not be applied to our data points 
of the classical method, which rather give additional systematics.

As shown in Figures \ref{fig:LF_thisstudy_compare_z5p7} and \ref{fig:LF_thisstudy_compare_z6p6}, 
the best-fit Schechter functions can explain the Ly$\alpha$ LF measurements 
in the wide luminosity range. 
If this is true, the faint-end slopes of Ly$\alpha$ LFs are very steep. 
The best-fit faint-end slope values are $\alpha = -2.5 - -2.6$ (Table \ref{table:LF_schechter}), 
which may indicate that the faint-end slopes of Ly$\alpha$ LFs 
are steeper than those of the UV LFs at similar redshifts 
(e.g., \cite{2015ApJ...803...34B}). 
Note that our best-fit faint-end slopes are steeper than 
that obtained in previous work on the $z=5.7$ Ly$\alpha$ LF \citep{2015ApJ...806...19D}.

It should be noted that, 
if we compare our Ly$\alpha$ LF measurements with 
the best-fit Schechter function results 
obtained from the classical method where we consider only the fainter Ly$\alpha$ luminosity range of 
$\log L_{\rm Ly\alpha}$ [erg s$^{-1}$] $\gtrsim 42.5-43.5$,  
we find that there is a significant bright-end excess of 
the $z = 5.7$ and $z = 6.6$ Ly$\alpha$ LF measurements 
at $\log L_{\rm Ly\alpha}$ [erg s$^{-1}$] $\gtrsim 43.5$. 
Based on the deviation of the bright-end data points from the best-fit Schechter function,
the significance value of the bright-end excesses is 
$\simeq 3 \sigma$ 
($2.6 \sigma$ for $z=5.7$ and $3.2 \sigma$ for $z=6.6$).
For $z = 6.6$, 
similar results are also claimed by some previous studies 
(e.g., \cite{2015MNRAS.451..400M}; \cite{2016MNRAS.463.1678S}; \cite{2016ApJ...818L...3C}; 
\cite{2017ApJ...837...11B}; \cite{2017ApJ...842L..22Z}).  
Although 
our 
results 
suggest that 
the LF fittings including the bright-end LF results 
may reveal the true shapes of the Ly$\alpha$ LFs, 
it is also possible that 
the bright-end LF results are enhanced by some systematic effects. 
We discuss possible origins of the bright-end excesses 
in Section \ref{sec:discuss_bright}.

\subsection{Comparison with Previous Studies}
\label{sec:LF_compare}

In this section, we compare our Ly$\alpha$ LFs at $z = 5.7$ and $6.6$
with those obtained by previous studies. 
As shown in Figures \ref{fig:LF_thisstudy_compare_z5p7} and \ref{fig:LF_thisstudy_compare_z6p6}, 
our Ly$\alpha$ LFs are generally consistent with those of the previous results. 
However, 
our Ly$\alpha$ LF results do not agree with 
the high number densities of LAEs 
recently claimed by 
\citet{2015MNRAS.451..400M} and \citet{2016MNRAS.463.1678S}.  
The reason of this discrepancy is unclear. 
This study and most of the previous studies 
have derived the Ly$\alpha$ LFs by the classical method and/or by using Monte Carlo simulations  
that take account of the two systematic uncertainties (I) and (II) in Ly$\alpha$ LF estimates (Section \ref{sec:LF_z5p76p6}). 
\citet{2015MNRAS.451..400M} and \citet{2016MNRAS.463.1678S} also 
appear to have considered these two uncertainties; 
they have applied filter profile correction for Ly$\alpha$ flux estimates 
and taken into account the incompleteness of the \textit{NB}-excess color selection. 
One possible explanation for the discrepancy is that 
their corrections are redundant, 
and that the correction factors are overestimated.
In fact, 
in our end-to-end Monte Carlo simulations, 
we have adopted 
a Schechter functional form for Ly$\alpha$ LFs and a Gaussian for Ly$\alpha$ EW$_0$ probability distributions, 
and have determined their best-fit functions simultaneously 
based on $\chi^2$ fitting to the observed surface number densities 
and the $\textit{BB}-\textit{NB}$ color distributions (Section \ref{sec:LF_z5p76p6}). 
In other words, the two systematic uncertainties are considered at the same time in our simulations. 
This is because these two systematic effects are closely related to each other. 
On the other hand, 
it seems that 
\citet{2015MNRAS.451..400M} 
have estimated the effects of the two uncertainties \textit{separately} in their Sections 4.1 and 4.3 
(See also \cite{2016MNRAS.463.1678S}), which might cause overcorrections due to the redundancy. 
Another possibility is the difference of the Ly$\alpha$ EW$_0$ distributions.    
In our simulations, 
we have adopted a Gaussian Ly$\alpha$ EW$_0$ probability distribution 
(e.g., \cite{2006PASJ...58..313S}; \cite{2007ApJ...667...79G}; \cite{2008ApJS..176..301O}). 
On the other hand, 
\citet{2015MNRAS.451..400M} do not describe 
what functional form is used for the Ly$\alpha$ EW$_0$ distribution 
in their calculations of the filter profile correction estimates 
and the color selection incompleteness estimates  
(see also \cite{2016MNRAS.463.1678S}). 
For example, 
if they assume an EW$_0$ value that is significantly smaller than the typical value for LAEs, 
they would obtain too large correction factors 
and thus too large Ly$\alpha$ LF measurements.

\section{Discussion}
\label{sec:discuss}

\begin{figure*}
\begin{center}
\includegraphics[width=17cm]{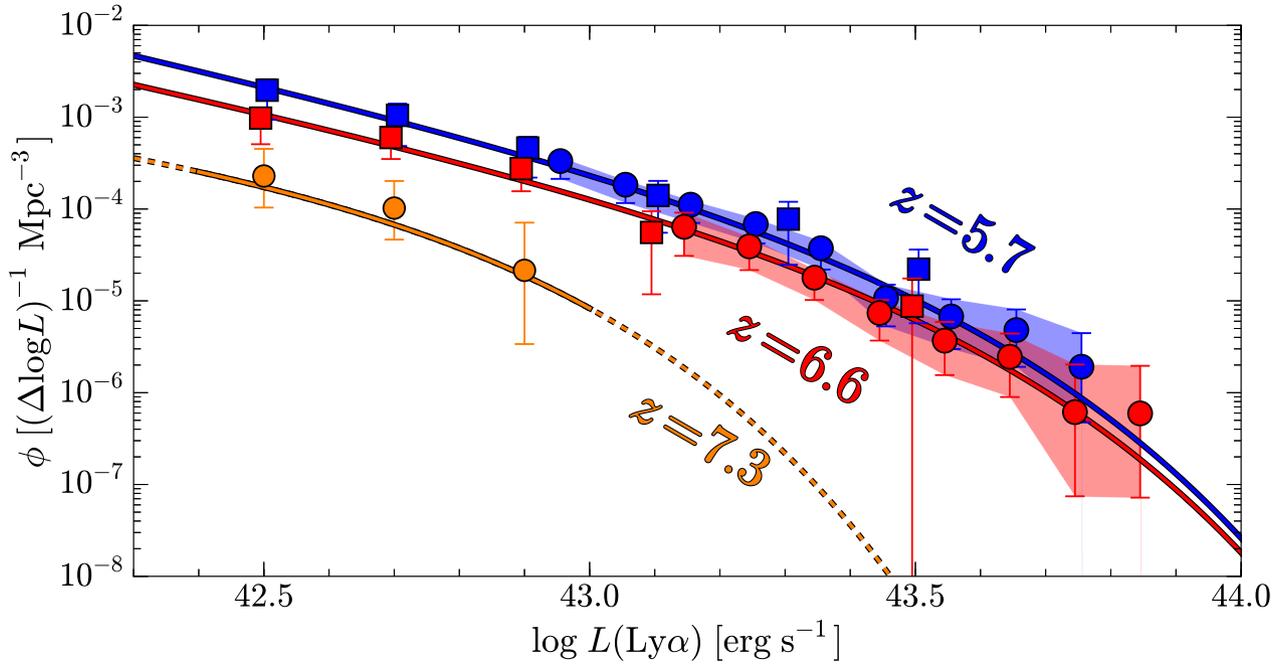}
\end{center}
\caption{
Evolution of the Ly$\alpha$ LFs.
The blue and red filled circles are our $z = 5.7$ and $6.6$ Ly$\alpha$ LF measurements, respectively,
which are derived from the HSC SSP data.
The blue and red filled squares denote the $z = 5.7$ and $6.6$ Ly$\alpha$ LF measurements
with the Subaru/Suprime-Cam data given by 
\citet{2008ApJS..176..301O} and \citet{2010ApJ...723..869O}, 
respectively.
The best-fit Schechter function for the Ly$\alpha$ LF at $z = 5.7$ ($6.6$)
is shown with the blue (red) solid curve, 
which are derived for the luminosity range of 
$\log L(\mathrm{Ly}\alpha)$ [erg s$^{-1}$] $= 42.4-44.0$. 
The orange circles represent the Ly$\alpha$ LF at $z = 7.3$ derived by 
\citet{2014ApJ...797...16K}. 
}\label{fig:LyaLF_evo_HSC}
\end{figure*}

\subsection{Systematic Effects in the Ly$\alpha$ LF Measurements} 
\label{sec:discuss_bright}

As shown in Section \ref{sec:LF_z5p76p6}, 
our best-fit Schechter functions 
derived with the end-to-end Monte Carlo simulations 
as well as the ones derived with the classical method 
for the Ly$\alpha$ luminosity range of 
$\log L_{\rm Ly\alpha}$ [erg s$^{-1}$] $\gtrsim 42.5-44.0$ 
are fitted to the Ly$\alpha$ LF measurements well 
both at the bright end and fainter magnitude bins. 
However, 
the best-fit values of the faint-end slope $\alpha$ are very steep,  
compared to 
the shallower slopes of the UV LFs at similar redshifts 
(e.g., \cite{2015ApJ...803...34B}).
Although our results may imply that 
the wide luminosity range of our Ly$\alpha$ LFs 
allow us to reveal the true shapes of the Ly$\alpha$ LFs, 
it is also possible that 
the bright-end measurements have some systematic effects. 
There are four possibilities for such systematics. 
One possibility is the contribution of AGNs, which is the same as the origin of the bright-end excess 
at $z \sim 2 - 3$ (e.g., \cite{2016ApJ...823...20K}). 
Another possibility is the formation of large ionized bubbles in the IGM around bright LAEs
during the epoch of reionization (EoR; e.g., \cite{2016MNRAS.463.1678S}; \cite{2017ApJ...837...11B}; \cite{2017ApJ...842L..22Z}). 
The possibility of the gravitational lensing effect
also needs to be considered 
(e.g., \cite{2011Natur.469..181W}; \cite{2011ApJ...742...15T}; \cite{2015ApJ...805...79M}). 
The other possibility is that 
merger systems which are blended at ground-based resolution appear as very bright LAEs 
(e.g., \cite{2017MNRAS.466.3612B}).

Firstly, we discuss the possibility of AGNs.
Although the number densities of AGNs rapidly decrease from $z \sim 3$
toward higher redshift 
(e.g., \cite{2012ApJ...746..125H}), 
some previous studies suggest the existence of (faint) AGNs at $z \sim 6-7$
(e.g., \cite{2010AJ....139..906W}; \cite{2011Natur.474..616M}; \cite{2015ApJ...798...28K}; 
\cite{2015A&A...578A..83G}; \cite{2016ApJ...833..222J}; \cite{2017MNRAS.469..448B}; 
\cite{2017arXiv170407750P} 
),  
which may 
systematically enhance the bright end of our $z = 5.7$ and $6.6$  Ly$\alpha$ LFs. 
To evaluate this possibility quantitatively,
we compare the number densities of faint AGNs presented in the literature 
with those of bright-end LAEs with $\log L_{\mathrm{Ly}\alpha}$ [erg s$^{-1}$] $> 43.5$. 
The numbers of bright-end LAEs at $z = 5.7$ and $6.6$ are 10 and 13, 
respectively.
Dividing the numbers of bright-end LAEs by the survey volumes (Section \ref{sec:sample_image}), 
we obtain their number densities of $8.6 \times 10^{-7}$ Mpc$^{-3}$ and $6.8 \times 10^{-7}$ Mpc$^{-3}$
at $z = 5.7$ and $6.6$, respectively.
Since the UV magnitudes of the bright-end LAEs are $M_\mathrm{UV} \gtrsim -21$ mag, 
we compare their number densities with 
extrapolations of the previous QSO UV LF results for brighter magnitudes 
(e.g., \cite{2010AJ....139..906W}; \cite{2015ApJ...798...28K}; \cite{2016ApJ...833..222J}). 
We find that 
the number densities of bright-end LAEs are consistent with the QSO UV LF results at $z \sim 6$, 
which indicates that bright-end LAEs with $\log L_{\mathrm{Ly}\alpha}$ [erg s$^{-1}$] $\gtrsim 43.5$ 
at $z = 5.7$ and $6.6$ could be AGNs. 
It should be noted that 
our recent deep near-infrared spectroscopic follow-up observations 
for several bright-end LAEs at $z = 5.7$ and $6.6$ 
reveal no clear signature of AGNs such as a broad Ly$\alpha$ emission line 
and strong highly-ionized metal lines, e.g., \textsc{Nv} and \textsc{Civ} 
\citep{2017arXiv170500733S}. 
Although these spectroscopy results imply that 
the observed bright-end LAEs are unlikely to host an AGN, 
the number of spectroscopically observed bright-end LAEs is still small. 
To further examine the possibility of AGNs, 
we will continue to carry out deep follow-up near-infrared spectroscopy.

Secondly, we discuss the possibility of large ionized bubbles.
During the EoR, Ly$\alpha$ photons can easily escape into the IGM 
in the case that the galaxy is surrounded by an ionized bubble which is 
large enough to allow the Ly$\alpha$ photons to redshift 
out of resonant scattering before entering the IGM at the edge of the ionized bubble 
(e.g., \cite{2015MNRAS.451..400M}; \cite{2017ApJ...837...11B}). 
In this case, 
it is expected that bright-end LAEs are preferentially observed, 
which 
can enhance the number densities of LAEs at the bright end. 
In other words, 
the $z=6.6$ bright-end LF may be enhanced 
by the effect of large ionized bubbles to some extent, 
although this effect is unlikely to happen at $z = 5.7$, 
where the IGM is already highly ionized (e.g., \cite{2006AJ....132..117F}). 
We further consider this possibility speculatively. 
By using the analytic models of 
\citet{2006MNRAS.365.1012F} (See also \cite{2005MNRAS.363.1031F}), 
we quantify the typical size of ionized bubbles around LAEs at $z = 6.6$.  
We use their results of the relations between the globally averaged ionized fraction of the IGM 
and the typical size of ionized bubbles, where overlaps of ionized bubbles are considered.   
As we will describe in Section \ref{sec:discuss_xHI},
we estimate the neutral hydrogen fraction at $z = 6.6$ to be 
$x_\mathrm{HI} = 0.3 \pm 0.2$  
from the evolution of the Ly$\alpha$ LFs at $z = 5.7 - 6.6$.
Based on the $x_\mathrm{HI}$ value and the top panel of Figure 1 of 
\citet{2006MNRAS.365.1012F}, 
we obtain the typical size of ionized bubbles at $z = 6.6$ of $\sim 15$ comoving Mpc.
If 
the bright-end excess at $z=6.6$ 
is caused by large ionized bubbles, 
the sizes of ionized bubbles around bright-end LAEs would be larger than $\sim 15$ comoving Mpc. 
To estimate the sizes of ionized bubbles around bright-end LAEs, 
we use the following formula for 
the Str{\"o}mgren radius $R_\mathrm{S}$ of an ionized bubble around a source at $z = 6.6$ by \citet{2002ApJ...576L...1H}:  
$R_\mathrm{S} = 0.8 \times (\mathrm{SFR}/10 \ M_\odot \ \mathrm{yr}^{-1})^{1/3} (t_{*}/100 \ \mathrm{Myr})^{1/3} [(1+z_{*})/7.56]^{-1}$ proper Mpc. 
In this equation, \citet{2002ApJ...576L...1H} 
considers an ionizing source at a given redshift $z_{*}$
with a constant SFR and a Salpeter IMF (the $0.1 - 120 M_\odot$ mass range), 
assuming that the source produces ionizing photons during the lifetime ($t_{*}$). 
From this equation and the UV magnitudes of the bright-end LAEs at $z = 6.6$ (i.e., $M_\mathrm{UV} \gtrsim -21$ mag),
we calculate the size of the ionized bubbles of $R_\mathrm{S} \lesssim 7$ comoving Mpc.\footnote{
The SFRs can be estimated from UV luminosities with the following equation: 
$\mathrm{SFR} \ (M_\odot \ \mathrm{yr}^{-1}) = L_\mathrm{UV} \ (\mathrm{erg} \ \mathrm{s}^{-1} \ \mathrm{Hz}^{-1})/(8 \times 10^{27})$ 
\citep{1998ApJ...498..106M}. 
From this equation,
the SFR corresponding to $M_\mathrm{UV} = -21$ is $13.6 \ M_\odot \ \mathrm{yr}^{-1}$.
We estimate the ionized bubble size under the assumption that
that these bright LAEs have a constant SFR of $13.6 \ M_\odot \ \mathrm{yr}^{-1}$,
and emit ionizing photons during their age of $100$ Myr.
}
This size is smaller than that estimated from the analytic model of 
\citet{2006MNRAS.365.1012F} 
($\sim 15$ comoving Mpc).
This result implies that, 
if 
the bright end of the Ly$\alpha$ LF at $z=6.6$ is 
enhanced by large ionized bubbles, 
ionizing sources that are different from the bright LAEs would be clustered 
around bright LAEs 
and form large ionized regions by overlapping their ionized bubbles.

Thirdly, we discuss the possibility of the gravitational lensing effect.
The lensing effect by foreground massive galaxies boosts 
apparent magnitudes of LAEs, which can make a bright-end excess of LFs 
(\cite{2011Natur.469..181W}; \cite{2011ApJ...742...15T}; \cite{2015ApJ...805...79M}; \cite{2015MNRAS.450.1224B}). 
To investigate whether the bright-end LAEs are affected by the gravitational lensing, 
we identify foreground sources around them which can act as lenses. 
We check a catalog of massive galaxy clusters 
that have been found 
by using the Cluster finding Algorithm based on Multi-band Identification of Red-sequence gAlaxies 
(CAMIRA; \cite{2014MNRAS.444..147O}; \cite{2017arXiv170100818O}).   
In addition, 
we check the positions of 
massive ($M_{\rm star} > 10^{10.3} M_\odot$) red galaxies with photometric redshift of $z_{\rm photo} = 0.05-1.05$ 
(M. Oguri et al. in preparation). 
However, 
we find that 
out of the 23 bright-end LAEs only two have a nearby foreground galaxy on the sky, 
which may produce modest lensing magnifications of $\mu\sim 1.2-1.7$.  
Thus, we conclude that 
the impact of the gravitational lensing on the shapes of the Ly$\alpha$ LFs is small.

Finally, we discuss the possibility of blended merging galaxies. 
Recently, \citet{2017MNRAS.466.3612B} 
have found that 
multi-component systems account for more than $40${\%} of 
their bright $z\sim7$ galaxies 
based on the analyses of their \textit{Hubble} images.  
In fact, our bright-end LAEs include well-studied Himiko and CR7, 
whose morphologies in the \textit{Hubble} WFC3 images show possible signatures of galaxy mergers 
(\cite{2013ApJ...778..102O}; \cite{2015ApJ...808..139S}). 
At least 
we confirm that the light profiles of our bright-end LAEs in the HSC images 
are mostly consistent with point sources 
\citep{2017arXiv170408140S}. 
However, 
the relatively coarse ground-based resolution cannot rule out the possibility that 
they are merging systems. 
To examine this possibility, 
we plan to investigate the morphologies of bright-end LAEs 
with higher resolution images taken with \textit{Hubble}.

In summary, 
the bright end of our Ly$\alpha$ LFs could be systematically enhanced by 
the contribution of AGNs and/or blended merging galaxies. 
It may also be possible that 
large ionized bubbles contribute to 
the bright end at $z=6.6$ 
if ionizing sources are clustered around bright-end LAEs.   
To further investigate the remaining possibilities, follow-up observations are needed.

\subsection{Evolution of Ly$\alpha$ LF at $z = 5.7 - 6.6$}
\label{sec:discuss_LFevo}

\begin{figure}
\begin{center}
\includegraphics[width=8cm]{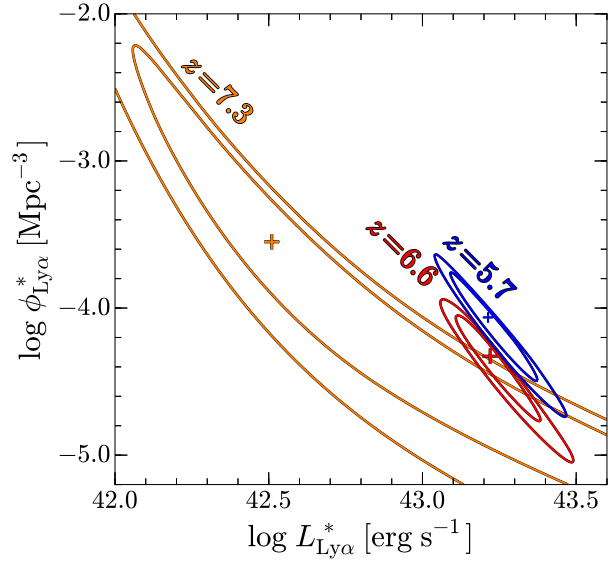}
\end{center}
\caption{
$1\sigma$ and $2\sigma$ confidence intervals of the Schechter parameters, 
$L^{*}_{\mathrm{Ly}\alpha}$ and $\phi^{*}_{\mathrm{Ly}\alpha}$.
The blue (red) 
contours correspond to $z = 5.7$ ($6.6$).
The blue and red crosses are the best-fit Schechter parameters
for the Ly$\alpha$ LFs at $z = 5.7$ and $6.6$, respectively.
These results are obtained 
with the classical method 
for the luminosity range of $\log L(\mathrm{Ly}\alpha)$ [erg s$^{-1}$] $= 42.4-44.0$. 
The orange contours show the results for the $z = 7.3$ Ly$\alpha$ LF 
\citep{2014ApJ...797...16K}. 
}\label{fig:error_cont}
\end{figure}

We investigate the evolution of the Ly$\alpha$ LF at $z = 5.7 - 6.6$.
In Figure \ref{fig:LyaLF_evo_HSC}, we show our Ly$\alpha$ LFs at $z = 5.7$ and $6.6$,
which are obtained from the $13.8$ deg$^2$ and $21.2$ deg$^2$ sky area of 
the HSC SSP survey. 
Here, we show 
the best-fit Schechter functions for the LF data points 
in the luminosity range of 
$\log L_{\rm Ly\alpha}$ [erg s$^{-1}$] $\gtrsim 42.4-44.0$ 
derived with the classical method, 
which are good approximations to the true LFs (Section \ref{sec:LF_z5p76p6}). 
We also present the Ly$\alpha$ LF at $z = 7.3$ derived by 
\citet{2014ApJ...797...16K} 
in this figure,
who have conducted the ultradeep $z = 7.3$ LAE survey with Subaru/Suprime-Cam.
\citet{2008ApJS..176..301O} and \citet{2010ApJ...723..869O} 
have derived the Ly$\alpha$ LFs at $z = 5.7$ and $6.6$
based on their $\sim 1$ deg$^2$ narrowband imaging data taken with Subaru/Suprime-Cam,
and have found the decrease of the Ly$\alpha$ LF from $z = 5.7$ to $6.6$.
The same results have been obtained by other previous studies
(e.g., \cite{2006ApJ...648....7K}; \cite{2010ApJ...725..394H}; \cite{2011ApJ...734..119K}; \cite{2016MNRAS.463.1678S}). 
We find such evolution from our Ly$\alpha$ LFs at $z = 5.7$ and $6.6$
in Figure \ref{fig:LyaLF_evo_HSC}.
To evaluate this evolution at $z = 5.7 - 6.6$ quantitatively,
we investigate the error distribution of Schechter parameters.
Figure \ref{fig:error_cont} presents the error contours of the Schechter parameters, 
$L^{*}_{\mathrm{Ly}\alpha}$ and $\phi^{*}_{\mathrm{Ly}\alpha}$, of
our $z = 5.7$ and $6.6$ Ly$\alpha$ LFs 
shown with the blue and red ovals, respectively.
We also show the error contours for the Ly$\alpha$ LF at $z = 7.3$ of 
\citet{2014ApJ...797...16K}. 
From this figure, the Schechter parameters of the $z = 6.6$ Ly$\alpha$ LF
are different from those of the $z = 5.7$ Ly$\alpha$ LF, 
and the Ly$\alpha$ LF decreases from $z = 5.7$ to $6.6$ at the $> 90 \%$ confidence level.
Note that 
the evolution of the Ly$\alpha$ LFs that we derive 
is similar to the one reported by \citet{2016MNRAS.463.1678S}, 
although our best-fit $L^{*}_{\mathrm{Ly}\alpha}$ values are smaller than theirs. 
The decreasing trend of the Ly$\alpha$ LFs with increasing redshift 
obtained in this study 
is also consistent with those of \citet{2010ApJ...723..869O}, 
who have investigated the evolution of LFs in the faint Ly$\alpha$ range 
($\log L(\mathrm{Ly}\alpha)$ [erg s$^{-1}$] $ \lesssim 43$) 
as shown in Figure \ref{fig:LyaLF_evo_HSC}. 
It should be noted that 
the best-fit Ly$\alpha$ LF parameters 
of $\phi^{*}_{\mathrm{Ly}\alpha}$ and $L^{*}_{\mathrm{Ly}\alpha}$ 
presented in Figure \ref{fig:error_cont} 
appear to be shifted from those of \citet{2010ApJ...723..869O}.  
This is caused by the difference of the faint-end slope $\alpha$ values.   
In our Schechter function fitting with the classical method, 
the slope $\alpha$ is treated as a free parameter 
and the best-fit value is about $-2.5$. 
On the other hand, 
in \citet{2010ApJ...723..869O} 
the $\alpha$ value has been fixed at $-1.5$.

\subsection{Estimation of $x_\mathrm{HI}$ at $z = 6.6$}
\label{sec:discuss_xHI}

We estimate the neutral hydrogen fraction, $x_\mathrm{HI}$, at $z = 6.6$
based on our Ly$\alpha$ LFs at $z = 5.7$ and $6.6$ 
in the same manner as 
\citet{2010ApJ...723..869O} and \citet{2014ApJ...797...16K}. 
We first calculate 
$T^{\mathrm{IGM}}_{\mathrm{Ly}\alpha, z = 6.6} / T^{\mathrm{IGM}}_{\mathrm{Ly}\alpha, z = 5.7}$,
where $T^{\mathrm{IGM}}_{\mathrm{Ly}\alpha, z}$ is a Ly$\alpha$ transmission
through the IGM at a redshift $z$.
The observed Ly$\alpha$ LD, $\rho^{\mathrm{Ly}\alpha}$, can be obtained from 
\begin{equation}
\rho^{\mathrm{Ly}\alpha} = \kappa \ T^{\mathrm{IGM}}_{\mathrm{Ly}\alpha, z} \ f^{\mathrm{esc}}_{\mathrm{Ly}\alpha} \ \rho^{\mathrm{UV}}, \label{eq:eq1}
\end{equation}
where $ \kappa$ is the conversion factor from UV to Ly$\alpha$ fluxes, $f^{\mathrm{esc}}_{\mathrm{Ly}\alpha}$
is the Ly$\alpha$ escape fraction through the ISM of a galaxy, and $\rho^{\mathrm{UV}}$ is the intrinsic UV LD.
Based on the equation, we can estimate 
the Ly$\alpha$ transmission fraction 
$T^{\mathrm{IGM}}_{\mathrm{Ly}\alpha, z = 6.6} / T^{\mathrm{IGM}}_{\mathrm{Ly}\alpha, z = 5.7}$
by
\begin{equation}
\frac{T^{\mathrm{IGM}}_{\mathrm{Ly}\alpha, z = 6.6}}{T^{\mathrm{IGM}}_{\mathrm{Ly}\alpha, z = 5.7}} = \frac{\kappa_{z = 5.7}}{\kappa_{z = 6.6}} \frac{f^{\mathrm{esc}}_{\mathrm{Ly}\alpha, z = 5.7}}{f^{\mathrm{esc}}_{\mathrm{Ly}\alpha, z = 6.6}} \frac{\rho^{\mathrm{Ly}\alpha}_{z = 6.6} / \rho^{\mathrm{Ly}\alpha}_{z = 5.7}}{\rho^{\mathrm{UV}}_{z = 6.6} / \rho^{\mathrm{UV}}_{z = 5.7}}.	\label{eq:eq2}
\end{equation}
To calculate $\rho^{\mathrm{Ly}\alpha, \mathrm{tot}}_{z = 6.6} / \rho^{\mathrm{Ly}\alpha, \mathrm{tot}}_{z = 5.7}$, 
we use the Ly$\alpha$ LD results in Section \ref{sec:LF_z5p76p6}. 
We adopt the Ly$\alpha$ LDs derived for the Ly$\alpha$ LF measurements in the luminosity range of  $\log L(\mathrm{Ly}\alpha)$ [erg s$^{-1}$] $= 42.4-44.0$, 
to take account of the contribution from bright-end LAEs 
as well as from the fainter ones. 
Based on the UV LF measurements of \citet{2015ApJ...803...34B}, 
$\rho^{\mathrm{UV}}_{z = 6.6} / \rho^{\mathrm{UV}}_{z = 5.7} = 0.74 \pm 0.10$ is obtained. 
Under the assumption of $\kappa_{z = 5.7}/\kappa_{z = 6.6} = 1$ 
and $f^{\mathrm{esc}}_{\mathrm{Ly}\alpha, z = 5.7} / f^{\mathrm{esc}}_{\mathrm{Ly}\alpha, z = 6.6} = 1$,
we obtain 
$T^{\mathrm{IGM}}_{\mathrm{Ly}\alpha, z = 6.6} / T^{\mathrm{IGM}}_{\mathrm{Ly}\alpha, z = 5.7} = 0.70 \pm 0.15$.
from Equation (\ref{eq:eq2}).

We obtain constraints on $x_\mathrm{HI}$ 
based on comparisons of our results with theoretical models. 
\citet{2004MNRAS.349.1137S} have calculated the IGM Ly$\alpha$ transmission fraction 
as a function of $x_\mathrm{HI}$ in two cases of galactic outflow: 
the Ly$\alpha$ velocity shifts of 0 and 360 km s$^{-1}$ from the systemic velocity. 
It is noted from recent studies that 
the average velocity shift of Ly$\alpha$ emission 
is $\sim 200$ km s$^{-1}$ for LAEs at $z \sim 2$
(e.g., \cite{2013ApJ...765...70H,2014ApJ...788...74S}). 
Based on Figure 25 of \citet{2004MNRAS.349.1137S}, 
our Ly$\alpha$ transmission fraction result 
is consistent with $x_\mathrm{HI}$ $\sim 0.0-0.2$ considering the two cases.
Next, we compare our Ly$\alpha$ LF result with 
the theoretical results of \citet{2007MNRAS.381...75M}, 
who have derived $z = 6.6$ Ly$\alpha$ LFs for various $x_\mathrm{HI}$ values 
based on their radiative transfer simulations.  
From Figure 4 of \citet{2007MNRAS.381...75M}, 
we obtain constraints of $x_\mathrm{HI} \sim 0.3-0.5$.  
Finally, 
we compare our result with a combination of two theoretical models. 
\citet{2007MNRAS.379..253D} have derived expected Ly$\alpha$ transmission fractions 
of the IGM as a function of the typical size of ionized bubbles 
(see also \cite{2007MNRAS.377.1175D}). 
The relation between the typical size of ionized bubbles and $x_\mathrm{HI}$ 
has been calculated by \citet{2006MNRAS.365.1012F} based on their analytic model.
A comparison of our Ly$\alpha$ transmission fraction result 
with these two models 
(Figure 6 of \cite{2007MNRAS.379..253D} and the top panel of Figure 1 of \cite{2006MNRAS.365.1012F})  
yields $x_\mathrm{HI} \sim 0.1-0.3$.  
Based on the results described above, 
we conclude the neutral hydrogen fraction is estimated to be 
$x_\mathrm{HI} = 0.1-0.5$, i.e., $x_\mathrm{HI} = 0.3 \pm 0.2$ 
at $z = 6.6$, 
where 
the variance of the theoretical model predictions 
as well as the uncertainties in our Ly$\alpha$ transmission fraction estimates 
are considered.

\begin{figure}
\begin{center}
\includegraphics[width=8cm]{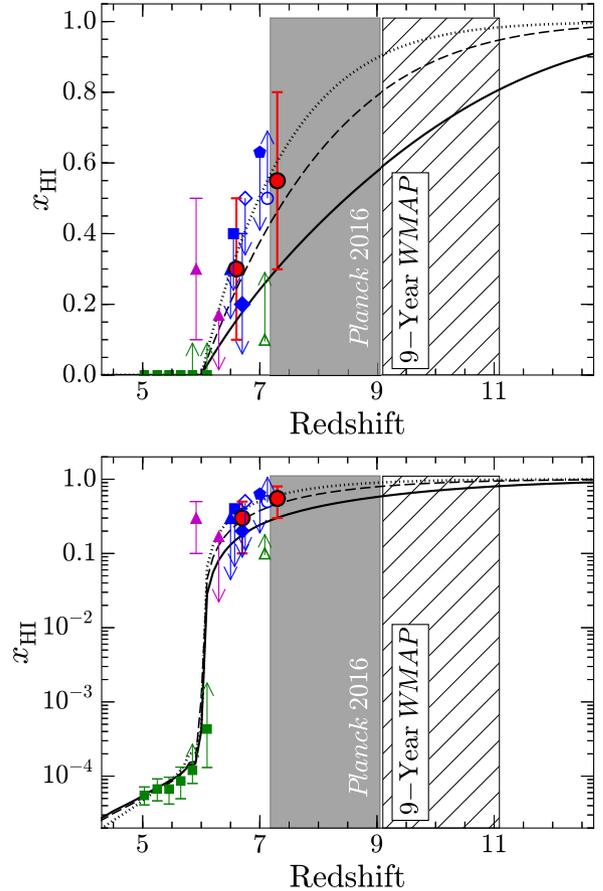}
\end{center}
\caption{
Evolution of the IGM neutral hydrogen fraction. 
The top and bottom panels are the same, 
except the scales of the ordinate axes (top: linear; bottom: logarithmic). 
The red filled circles show the $x_\mathrm{HI}$ estimates from the Ly$\alpha$ LFs at $z = 6.6$ 
(this study) and $7.3$ 
\citep{2014ApJ...797...16K}. 
The blue filled triangle, square, diamond, and pentagon are 
the $x_\mathrm{HI}$ estimates based on the evolution of the Ly$\alpha$ LF 
obtained by \citet{2004ApJ...617L...5M}, \citet{2011ApJ...734..119K}, \citet{2010ApJ...723..869O}, 
and \citet{2010ApJ...722..803O}, respectively.
The blue open diamond and circle are 
the constraints on $x_\mathrm{HI}$ from the clustering analyses of LAEs 
\citep{2010ApJ...723..869O} 
and the Ly$\alpha$ emitting galaxy fraction 
(\cite{2011ApJ...743..132P}; \cite{2012ApJ...744..179S}; \cite{2012ApJ...744...83O}; 
\cite{2012ApJ...747...27T}; \cite{2012MNRAS.427.3055C}; \cite{2014MNRAS.443.2831C}; 
\cite{2014ApJ...793..113P}; \cite{2014ApJ...795...20S}), respectively.
The previous results from the GRB optical afterglow spectrum analyses 
are shown with magenta filled triangles 
\citep{2006PASJ...58..485T,2014PASJ...66...63T}. 
The green filled squares and open triangle are 
the results from the GP test of QSOs 
\citep{2006AJ....132..117F}
and the size of QSO near zones 
\citep{2011Natur.474..616M,2011MNRAS.416L..70B},
respectively.
The gray and hatched regions are 
the $1\sigma$ confidence intervals 
for the instantaneous reionization redshifts obtained by 
\textit{Planck} \citep{2016A&A...596A.107P} 
and 
nine-year \textit{WMAP} \citep{2013ApJS..208...19H,2013ApJS..208...20B}, 
respectively.
The models A, B, and C of \citet{2008MNRAS.385L..58C}
are shown with doted, dashed and solid lines, respectively. 
}\label{fig:xHI_evolution}
\end{figure}

Figure \ref{fig:xHI_evolution} shows our $x_\mathrm{HI}$ estimate at $z = 6.6$ 
and those taken from the previous studies. 
The previous results of the $z \gtrsim 7$ Ly$\alpha$ LFs imply $x_\mathrm{HI} = 0.3 - 0.8$ at $z = 7.3$ 
\citep{2014ApJ...797...16K} 
and $x_\mathrm{HI} < 0.63$ at $z = 7.0$ 
\citep{2010ApJ...722..803O}. 
The studies of Ly$\alpha$ emitting fractions indicate $x_\mathrm{HI} \gtrsim 0.5$ at $z \sim 7$
(e.g., \cite{2011ApJ...743..132P}; \cite{2012ApJ...744..179S}; \cite{2012ApJ...744...83O}; 
\cite{2012ApJ...747...27T}; \cite{2012MNRAS.427.3055C}; \cite{2014MNRAS.443.2831C}; 
\cite{2014ApJ...793..113P}; \cite{2014ApJ...795...20S}). 
The Ly$\alpha$ damping wing absorption measurements of QSOs 
suggest $x_\mathrm{HI} \gtrsim 0.1$ at $z = 7.1$ 
\citep{2011Natur.474..616M,2011MNRAS.416L..70B}.

\begin{figure}
\begin{center}
\includegraphics[width=8cm]{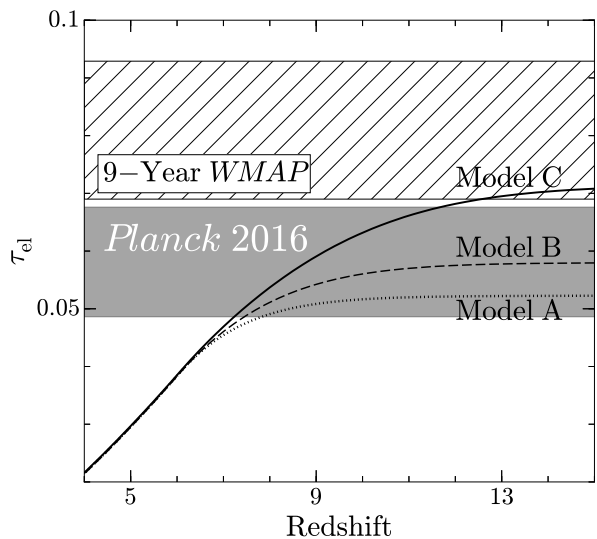}
\end{center}
\caption{Thomson scattering optical depth, $\tau_\mathrm{el}$, 
as a function of redshift.
The gray and hatched regions correspond to 
the $1\sigma$ confidence intervals of the $\tau_\mathrm{el}$ measurements
obtained by 
the \textit{Planck} 2016 data \citep{2016A&A...596A.107P} 
and 
the nine-year \textit{WMAP} data \citep{2013ApJS..208...19H,2013ApJS..208...20B},  
respectively.
The doted, dashed, and solid curves are the models A, B, and C 
of \citet{2008MNRAS.385L..58C}, respectively. 
}\label{fig:tauel}
\end{figure}

As already pointed out in our previous work \citep{2014ApJ...797...16K}, 
the decrease of the Ly$\alpha$ LF from $z = 6.6$ to $7.3$ 
is larger than that from $z = 5.7$ to $6.6$.
In Figure \ref{fig:xHI_evolution}, 
this accelerated evolution could be also found, although the uncertainties are large. 
The Ly$\alpha$ LF evolves from $z = 6.6$ to $7.3$ at the $>90${\%} confidence level,  
while 
the difference of $x_\mathrm{HI}$ between $z = 6.6$ and $7.3$ is only within $1 \sigma$. 
This is because, 
in our $x_\mathrm{HI}$ estimates, 
we take into account the uncertainties of  
the UV LFs and the various theoretical model results 
as well as the uncertainties of the Ly$\alpha$ LFs  
(see \cite{2014ApJ...797...16K} for details).

Here, we investigate whether the $x_\mathrm{HI}$ evolution obtained by our and previous studies
can explain the Thomson scattering optical depth, $\tau_\mathrm{el}$, value 
obtained from the latest \textit{Planck} 2016 data.
Because one needs to know $\tau_\mathrm{el}$ from a given $x_\mathrm{HI}$ evolution,
we use the semi-analytic models of \citet{2008MNRAS.385L..58C}. 
They have derived $x_\mathrm{HI}$ and $\tau_\mathrm{el}$ evolutions  
by considering three models 
which differ the minimum halo masses for reionization sources 
to cover typical scenarios of the cosmic reionization history. 
These three models are referred to as models A, B, and C 
corresponding to the minimum halo masses of 
$\sim 10^{9}$, $\sim 10^{8}$, and $\sim 5 \times 10^{5} \ M_{\odot}$, respectively, at $z = 6$. 
We present the $x_\mathrm{HI}$ evolutions of the three models in Figure \ref{fig:xHI_evolution},
and their $\tau_\mathrm{el}$ evolutions in Figure \ref{fig:tauel}.
The gray (hatched) region in Figure \ref{fig:tauel} 
shows the $1 \sigma$ range of $\tau_\mathrm{el}$ 
obtained by \textit{Planck} (\textit{WMAP}).
The latest results from the \textit{Planck} observations indicate that
the Thomson scattering optical depth 
is $\tau_\mathrm{el} = 0.058 \pm 0.009$ 
\citep{2016A&A...596A.107P}, 
which is significantly lower than the one obtained from the \textit{WMAP} data.
In Figure \ref{fig:xHI_evolution}, 
the models A and B are consistent with
our $x_\mathrm{HI}$ estimates at $z = 6.6$ and $7.3$,
and also explain the Thomson scattering optical depth
obtained by the latest \textit{Planck} 2016 data in Figure \ref{fig:tauel}.
The model C can barely explain our $x_\mathrm{HI}$ value at $z = 7.3$,
but is placed above the $\tau_\mathrm{el}$ of \textit{Planck} beyond the $1\sigma$ error
(Figure \ref{fig:tauel}).
Thus, these results show that the cosmic reionization history 
such as the models A and B 
can explain both the $x_\mathrm{HI}$ estimates and the \textit{Planck} 2016
$\tau_\mathrm{el}$ value simultaneously.
Similar conclusions are reached by 
\citet{2015ApJ...802L..19R} and \citet{2015ApJ...811..140B}, 
who have discussed the UV LF evolution of reionization sources that is independent from our Ly$\alpha$ LF study.

\section{Summay}
\label{sec: summary}

We have derived the Ly$\alpha$ LFs at $z = 5.7$ and $6.6$
based on the first-year narrowband and broadband imaging data products 
obtained by the HSC SSP survey.  
Our major results are listed below:

\begin{enumerate}

\item
Our HSC narrowband images for $z = 5.7$ and $6.6$ LAEs have the effective areas of  
$\sim 13.8$ deg$^2$ and $\sim 21.2$ deg$^2$, respectively.
The $5\sigma$ limiting magnitudes of the narrowband images are
$\sim 25.0$ mag and $\sim 25.5$ mag in the Deep and UltraDeep layers, respectively.
Using these narrowband images, we have identified, in total, $\sim$ 2,000 LAEs at $z = 5.7$ and $6.6$
with a bright Ly$\alpha$ luminosity range of $\log L(\mathrm{Ly}\alpha)$ [erg s$^{-1}$] $\simeq 42.9 - 43.8$.
Our HSC LAE sample is $\sim 2-6$ times larger than those of previous studies of $z \sim 6-7$ LAEs.

\item
Based on the LAE samples, we have derived the Ly$\alpha$ LFs at $z = 5.7$ and $6.6$.
We have obtained the best-fit Schechter parameters of 
$L^{*}_{\mathrm{Ly}\alpha} = 1.6^{+2.2}_{-0.6} \times 10^{43} \ \mathrm{erg} \ \mathrm{s}^{-1}$, 
$\phi^{*}_{\mathrm{Ly}\alpha} = 0.85^{+1.87}_{-0.77} \times 10^{-4} \ \mathrm{Mpc}^{-3}$, 
and 
$\alpha = -2.6^{+0.6}_{-0.4}$ 
for the $z = 5.7$ Ly$\alpha$ LF, 
and  
$L^{*}_{\mathrm{Ly}\alpha} = 1.7^{+0.3}_{-0.7} \times 10^{43} \ \mathrm{erg} \ \mathrm{s}^{-1}$, 
$\phi^{*}_{\mathrm{Ly}\alpha} = 0.47^{+1.44}_{-0.44} \times 10^{-4} \ \mathrm{Mpc}^{-3}$, 
and 
$\alpha = -2.5^{+0.5}_{-0.5}$  
for the $z = 6.6$ Ly$\alpha$ LF, 
if we consider the Ly$\alpha$ luminosity range of 
$\log L(\mathrm{Ly}\alpha)$ [erg s$^{-1}$] $= 42.4 - 44.0$.

\item
Our Ly$\alpha$ LFs at $z=5.7$ and $z=6.6$ 
show a very steep faint-end slope, 
although there is a possibility that 
the bright-end measurements are enhanced by some systematic effects 
such as 
the contribution from AGNs, 
blended merging galaxies, 
and/or 
large ionized bubbles around bright LAEs.

\item
We have confirmed the decrease of the Ly$\alpha$ LF from $z = 5.7$ to $6.6$.
This evolution is caused by the Ly$\alpha$ damping wing absorption of neutral hydrogen in the IGM.
Based on the decrease of the Ly$\alpha$ LF at $z = 5.7 - 6.6$,
we have estimated the IGM neutral hydrogen fraction of $x_\mathrm{HI} = 0.3 \pm 0.2$  at $z = 6.6$.
The $x_\mathrm{HI}$ evolution obtained from our and previous studies
can explain the Thomson scattering optical depth measurement of the latest \textit{Planck} 2016.
 
\end{enumerate}

\begin{ack}
We thank Mamoru Doi, 
Kentaro Motohara, Toshitaka Kajino, and Masafumi Ishigaki for useful discussion and comments.
We appreciate Masayuki Umemura and Masao Mori, who provided the fund for
the narrowband filters.

The Hyper Suprime-Cam (HSC) collaboration includes the astronomical communities
of Japan and Taiwan, and Princeton University.
The HSC instrumentation and software were developed by
the National Astronomical Observatory of Japan (NAOJ),
the Kavli Institute for the Physics and Mathematics of the Universe (Kavli IPMU),
the University of Tokyo, the High Energy Accelerator Research Organization (KEK),
the Academia Sinica Institute for Astronomy and Astrophysics in Taiwan (ASIAA),
and Princeton University.
Funding was contributed by the FIRST program from Japanese Cabinet Office,
the Ministry of Education, Culture, Sports, Science and Technology (MEXT),
the Japan Society for the Promotion of Science (JSPS), Japan Science and Technology Agency (JST),
the Toray Science Foundation, NAOJ, Kavli IPMU, KEK, ASIAA, and Princeton University. 

This paper makes use of software developed for the Large Synoptic Survey Telescope.
We thank the LSST Project for making their code available as free software at  http://dm.lsst.org

The Pan-STARRS1 Surveys (PS1) have been made possible through contributions
of the Institute for Astronomy, the University of Hawaii, the Pan-STARRS Project Office,
the Max-Planck Society and its participating institutes, the Max Planck Institute for Astronomy,
Heidelberg and the Max Planck Institute for Extraterrestrial Physics, Garching, The Johns Hopkins University,
Durham University, the University of Edinburgh, Queen's University Belfast,
the Harvard-Smithsonian Center for Astrophysics,
the Las Cumbres Observatory Global Telescope Network Incorporated,
the National Central University of Taiwan, the Space Telescope Science Institute,
the National Aeronautics and Space Administration under Grant No. NNX08AR22G issued
through the Planetary Science Division of the NASA Science Mission Directorate,
the National Science Foundation under Grant No. AST-1238877, the University of Maryland,
and Eotvos Lorand University (ELTE) and the Los Alamos National Laboratory.

Based on data collected at the Subaru Telescope and retrieved from the HSC data archive system,
which is operated by Subaru Telescope and Astronomy Data Center, NAOJ.

A.K. acknowledges support from the Japan Society for the Promotion of Science (JSPS)
through the JSPS Research Fellowship for Young Scientists. 
This work is supported by World Premier International Research
Center Initiative (WPI Initiative), MEXT, Japan, and
KAKENHI (15H02064) Grant-in-Aid for Scientific Research (A)
through Japan Society for the Promotion of Science. 
N.K. acknowledges supports from the JSPS grant 15H03645.

\end{ack}


\appendix 

\end{document}